%% file: main.tex
\documentclass[10pt,journal,compsoc]{IEEEtran}
\makeatletter\if@twocolumn\PassOptionsToPackage{switch}{lineno}\else\fi\makeatother
\input{_header}

\input{_data}

\usepackage{graphicx}
\usepackage{xurl}
\usepackage[T1]{fontenc}

\usepackage{lettrine}
\usepackage{booktabs}
\usepackage{xcolor}
\usepackage{balance}
\usepackage{enumitem}
\usepackage{listings}
\usepackage{colortbl}
\usepackage{threeparttable}
\usepackage{multirow}
\usepackage[framemethod=TikZ]{mdframed}

\usepackage[ruled,linesnumbered]{algorithm2e}

\renewcommand\thesection{\arabic{section}} 
\renewcommand\thesubsectiondis{\thesection.\arabic{subsection}}

\ifCLASSOPTIONcompsoc
  \usepackage[nocompress]{cite}
\else
  \usepackage{cite}
\fi
\ifCLASSINFOpdf
\else
\fi
\usepackage{amsmath}
\usepackage[cmintegrals]{newtxmath}
\usepackage{url,multirow,morefloats,floatflt,cancel,tfrupee}
\makeatletter

\AtBeginDocument{\@ifpackageloaded{textcomp}{}{\usepackage{textcomp}}}
\makeatother
\usepackage{colortbl}
\usepackage{xcolor}
\usepackage{pifont}
\usepackage[nointegrals]{wasysym}
\makeatletter

\def\mcWidth#1{\csname TY@F#1\endcsname+\tabcolsep}

\def\cAlignHack{\rightskip\@flushglue\leftskip\@flushglue\parindent\z@\parfillskip\z@skip}
\def\rAlignHack{\rightskip\z@skip\leftskip\@flushglue \parindent\z@\parfillskip\z@skip}

\@ifundefined{etal}{\def\etal{\textit{et~al}}}{}

\usepackage{ifxetex}
\ifxetex\else\if@twocolumn\@ifpackageloaded{stfloats}{}{\usepackage{dblfloatfix}}\fi\fi

\AtBeginDocument{
\expandafter\ifx\csname eqalign\endcsname\relax
\def\eqalign#1{\null\vcenter{\def\\{\cr}\openup\jot\m@th
  \ialign{\strut$\displaystyle{##}$\hfil&$\displaystyle{{}##}$\hfil
      \crcr#1\crcr}}\,}
\fi
}

\AtBeginDocument{%
  \@ifpackageloaded{endfloat}%
   {\renewcommand\efloat@iwrite[1]{\immediate\expandafter\protected@write\csname efloat@post#1\endcsname{}}}{\newif\ifefloat@tables}%
}%

\def\BreakURLText#1{\@tfor\brk@tempa:=#1\do{\brk@tempa\hskip0pt}}
\let\lt=<
\let\gt=>
\def\processVert{\ifmmode|\else\textbar\fi}

\@ifundefined{subparagraph}{
\def\subparagraph{\@startsection{paragraph}{5}{2\parindent}{0ex plus 0.1ex minus 0.1ex}%
{0ex}{\normalfont\small\itshape}}%
}{}

\newcommand\role[1]{\unskip}
\newcommand\aucollab[1]{\unskip}
  
\@ifundefined{tsGraphicsScaleX}{\gdef\tsGraphicsScaleX{1}}{}
\@ifundefined{tsGraphicsScaleY}{\gdef\tsGraphicsScaleY{.9}}{}
\def\checkGraphicsWidth{\ifdim\Gin@nat@width>\linewidth
	\tsGraphicsScaleX\linewidth\else\Gin@nat@width\fi}

\def\checkGraphicsHeight{\ifdim\Gin@nat@height>.9\textheight
	\tsGraphicsScaleY\textheight\else\Gin@nat@height\fi}

\def\fixFloatSize#1{}
\let\ts@includegraphics\includegraphics

\def\inlinegraphic[#1]#2{{\edef\@tempa{#1}\edef\baseline@shift{\ifx\@tempa\@empty0\else#1\fi}\edef\tempZ{\the\numexpr(\numexpr(\baseline@shift*\f@size/100))}\protect\raisebox{\tempZ pt}{\ts@includegraphics{#2}}}}

\AtBeginDocument{\def\includegraphics{\@ifnextchar[{\ts@includegraphics}{\ts@includegraphics[width=\checkGraphicsWidth,height=\checkGraphicsHeight,keepaspectratio]}}}

\DeclareMathAlphabet{\mathpzc}{OT1}{pzc}{m}{it}

\def\URL#1#2{\@ifundefined{href}{#2}{\href{#1}{#2}}}

\def\UrlOrds{\do\*\do\-\do\~\do\'\do\"\do\-}%
\g@addto@macro{\UrlBreaks}{\UrlOrds}

\edef\fntEncoding{\f@encoding}

\makeatother

\newif\ifmultipleabstract\multipleabstractfalse%
\makeatletter
\AtBeginDocument{\@ifpackageloaded{longtable}{%
\def\LT@makecaption#1#2#3{%
  \LT@mcol\LT@cols c{\hbox to\z@{\hss\parbox[t]\LTcapwidth{%
    \sbox\@tempboxa{#1{#2: } #3}%
    \ifdim\wd\@tempboxa>\hsize
      #1{#2: }\textsc{#3}%
    \else
      \hbox to\hsize{\hfil\box\@tempboxa\hfil}%
    \fi
    \endgraf\vskip\baselineskip}%
  \hss}}}
}{}}
\makeatother

\makeatletter
\let\citep\cite
\let\citet\cite
\makeatother

\begin{document}


%
\title{Leakage-Resilient and Carbon-Neutral Aggregation Featuring the Federated AI-enabled Critical Infrastructure}

%
%
\author{Zehang Deng, Ruoxi Sun,~\IEEEmembership{Member,~IEEE}, Minhui Xue,~\IEEEmembership{Member,~IEEE},\\ Sheng Wen,~\IEEEmembership{Senior Member,~IEEE}, Seyit Camtepe,~\IEEEmembership{Senior Member,~IEEE}, \\ Surya Nepal,~\IEEEmembership{Senior Member,~IEEE}, Yang Xiang,~\IEEEmembership{Fellow,~IEEE}




\thanks{Z. Deng, S. Wen and Y. Xiang are with School of Science, Computing and Engineering Technologies, Swinburne University of Technology, Melbourne, VIC Australia. Emails: \{zehangdeng,swen, yxiang\}@swin.edu.au}


\thanks{R. Sun, M. Xue, Seyit Camtepe and Surya Nepal are with the Cybersecurity and Quantum Systems Group, CSIRO’s Data61. Email:\{ruoxi.sun, jason.xue, seyit.camtepe, surya.nepal\}@data61.csiro.au}
\thanks{Corresponding Author: S. Wen}
}
\IEEEtitleabstractindextext{
\begin{abstract}
AI-enabled critical infrastructures (ACIs) integrate artificial intelligence (AI) technologies into various essential systems and services that are vital to the functioning of society, offering significant implications for efficiency, security and resilience. While adopting decentralized AI approaches (such as federated learning technology) in ACIs is plausible, private and sensitive data are still susceptible to data reconstruction attacks through gradient optimization. In this work, we propose Compressed Differentially Private Aggregation (CDPA), a leakage-resilient, communication-efficient, and carbon-neutral approach for ACI networks. Specifically, CDPA has introduced a novel random bit-flipping mechanism as its primary innovation. This mechanism first converts gradients into a specific binary representation and then selectively flips masked bits with a certain probability. The proposed bit-flipping introduces a larger variance to the noise while providing differentially private protection and commendable efforts in energy savings while applying vector quantization techniques within the context of federated learning. The experimental evaluation indicates that CDPA can reduce communication cost by half while preserving model utility. Moreover, we demonstrate that CDPA can effectively defend against state-of-the-art data reconstruction attacks in both computer vision and natural language processing tasks. We highlight existing benchmarks that generate 2.6x to over 100x more carbon emissions than CDPA. 
We hope that the CDPA developed in this paper can inform the federated AI-enabled critical infrastructure of a more balanced trade-off between utility and privacy, resilience protection, as well as a better carbon offset with less communication overhead. 
\end{abstract}
}

\maketitle

\IEEEdisplaynontitleabstractindextext

%
\IEEEpeerreviewmaketitle

\section{Introduction}

\begin{figure}[t]
\centering
\includegraphics[width=\linewidth]{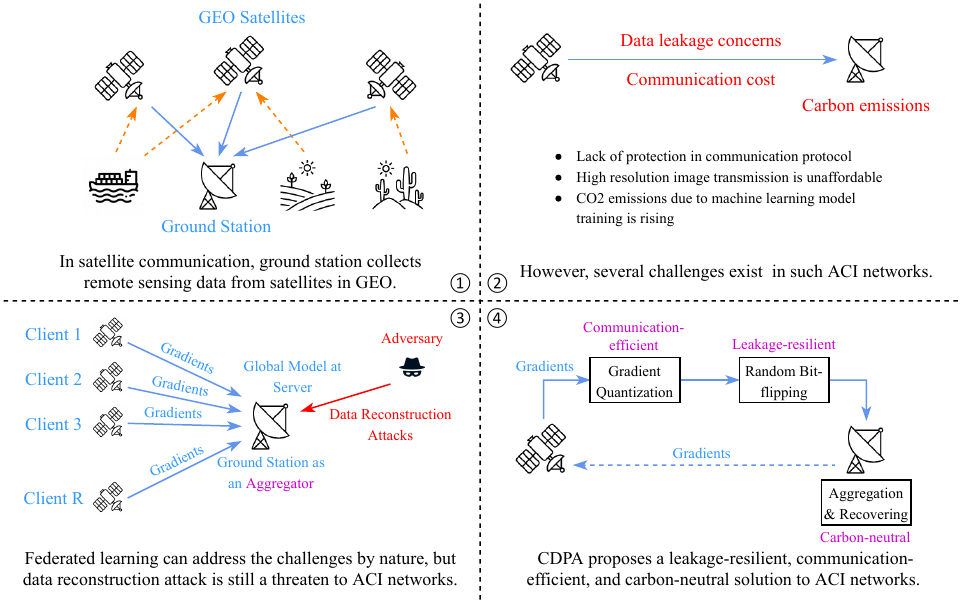}
\caption{An example of AI-enabled critical infrastructure. High-resolution data captured by remote sensing satellites in GEO pave the way for AI-driven advancements. Federated learning emerges as the solution of choice, as satellite communication struggles with image transmission and privacy protection.}
\label{fig_satellite_communication}
\end{figure}

Critical infrastructures are those essential facilities and services that are vital for the functioning of society and its economy, including energy systems, transportation networks, water supplies, satellite communications, and healthcare. The use of artificial intelligence (AI) in critical infrastructure systems has become increasingly significant in recent years~\cite{laplante2020artificial}. 
The integration of AI technologies into critical infrastructures involves the use of machine learning algorithms, data analytics, and automation to enhance the efficiency, reliability, and security of these systems. For example, AI optimizes the operation of smart grids by predicting and balancing energy supply and demand in real time, resulting in more efficient energy distribution and reduced power outages~\cite{omitaomu2021artificial,bose2017artificial,khan2023artificial}. 
Natural language processing (NLP) technology has been widely used in AI-enabled critical infrastructures (ACIs), such as maintenance and injury reports analysis in wind industry~\cite{gillen2017natural} and issue management in nuclear power plant~\cite{atomassist2023}.
In another emerging ACI scenario, as illustrated in Figure~\ref{fig_satellite_communication}, Geostationary Earth Orbit (GEO) satellites with AI can track environmental changes, such as agricultural monitoring, climate change analysis and air quality assessment, helping scientists and policymakers monitor climate change and take appropriate actions~\cite{fourati2021artificial,homssi2022artificial,chen2022satellite}.

The use of AI in critical infrastructures can be a two-headed coin. 
Recent research~\cite{graham2024navigating,oche2021applications} highlights the critical need for rigorous security risk analysis, particularly concerning vulnerabilities in satellite software, as well as concerns about confidentiality, accessibility, and legacy system threats across various space-domain use cases. Considering these insights, it is imperative to address AI-specific risks, such as data reconstruction attacks, and provide practical mitigation solutions tailored to the challenges in ACI scenarios.
\textbf{Data leakage concern} arises particularly when ACIs involve collecting and analyzing large amounts of data.
For example, the lack of encryption in the communication protocol and the performance overhead~\cite{huwyler2023qpep} lead to a trade-off between performance and data protection, exposing the ACI network to a plethora of attacks~\cite{border2001performance}. 
Meanwhile, \textbf{communication cost} of ACI networks could be another vital challenge.  
The newest GEO satellite, Hughes Jupiter 3, launched recently, dramatically expands the communication capacity to 500 Gbps with a cost over \$500 million~\cite{hughes2023}. 
However, this bandwidth must be shared among tens of thousands of potential Internet users simultaneously, resulting in expected bandwidth capping and throttling.
We note that communication cost concerns also exist in other ACIs, such as smart grids, AI-driven precision farming, and the oil and gas industry, as they are typically located in remote regions, away from large populations~\cite{tao2023transimitting}.
Furthermore, to achieve sustainable development, it is essential to reduce the \textbf{carbon emissions} of ACIs to withstand climate-related challenges~\cite{sachs2019six}. Recent research on the carbon footprint of AI argues that training AI models is highly polluting and encourages people to shift toward more efficient practices within the AI industry~\cite{kaack2022aligning,luccioni2023counting}. It is estimated that training a single transformer model can generate CO2 emissions roughly equal to the total lifetime carbon footprint of 5 cars~\cite{strubell2019energy}.

\input{Tables/tab_related_work}

Although Federated Learning (FL)~\cite{shokri_privacy-preserving_2015,konecny_federated_2016,zhang2021survey} has achieved great success in training models in a distributed manner, particularly in safeguarding individual user data from being collected by central servers, we anticipate that high communication and gradient optimization overhead will challenge the adoption of FL technology in ACIs when it requires transmitting gradient updates between clients and the server.
Moreover, the exposure of gradients or parameters during the transfer of data, as observed in \textit{data reconstruction attacks}~\cite{yin2021see,zhu2019deep,geiping2020inverting,yue2023gradient,deng2021tag,balunovic2022lamp}, continues to pose a threat to the confidentiality of the client's dataset, potentially resulting in \textbf{data leakage}.

With all the aforementioned challenges in mind, in this study, we propose \toollong (\tool) as a leakage-resilient, communication-efficient, and carbon-neutral federated aggregation solution towards featuring the AI-enabled critical infrastructure. 
Specifically, in \tool, to enhance ACIs' resilience to data leakage and protect sensitive data against state-of-the-art data reconstruction attacks (\eg DLG~\cite{zhu2019deep}, IG~\cite{geiping2020inverting}, ROG~\cite{yue2023gradient}, TAG~\cite{deng2021tag}, and LAMP~\cite{balunovic2022lamp} attacks), unlike conventional approaches such as Uveqfed (\ie SDQ) tailored mainly for compression, we design a random \textbf{bit-flipping} approach to add provable differentially private (DP) noise in data transmission. Different from existing DP methods, the proposed random bit-flipping introduces a larger variance in the added noise, thus enabling stronger protection. In addition, to address the challenges of communication cost, the tight real-time requirements of ACI applications, and the constrained hardware computational performance, we propose a flexible framework to integrate model compression techniques in ACIs. Particularly, \textit{\tool} incorporates Subtractive Dithered Lattice Quantization (SDQ)~\cite{shlezinger2020uveqfed} to curtail the communication cost while minimizing quantization error. We note that other model compression approaches such as gradient pruning~\cite{aji2017sparse}, compressive sensing~\cite{miao2022compressed}, quantization~\cite{bernstein2018signsgd,bernstein2018signsgd2,shlezinger2020uveqfed,alistarh2017qsgd}, and embedding representation~\cite{sriram2022deepcomp,theis2022lossy} could also be seamlessly integrated, but they may contribute differently to the final reconstruction error.
Furthermore, since the computational overhead of bit-flipping only entails converting gradients into a specific binary representation and directly flipping masked bits within this representation, 
it will not prolong the training time, thereby paving the way to carbon neutrality. 
To validate, we comprehensively evaluate its utility and defense performance in both computer vision (CV) and natural language processing (NLP) tasks, across different model architectures and various model sizes ranging from 9K to 110M parameters. We further compare \tool with several existing methods utilizing gradient compression (\eg  signSGD~\cite{bernstein2018signsgd,bernstein2018signsgd2}, GradDrop~\cite{aji2017sparse}, Soteria~\cite{sun2021soteria} and SDQ~\cite{shlezinger2020uveqfed}) and differential privacy (\eg LDP~\cite{dwork2014algorithmic}, JoPEQ~\cite{lang2023joint} and CAFL~\cite{miao2022compressed}).
The experimental results indicate that \tool not only preserves the model's utility but also demonstrates robustness of defenses in both CV and NLP tasks. Particularly, \tool achieves LPIPS 0.732 in CV tasks and reduces the reconstructed word count to 0 in NLP tasks without sacrificing model utility. 
We also demonstrate that practical carbon emission is significantly reduced, compared to existing approaches. The source code is publicly available at \url{https://github.com/CDPAdefense/CDPA.git}.

In summary, our main contributions are as follows.

\begin{itemize}[leftmargin=*]
\item We propose \tool, a novel federated aggregation approach with a provable privacy guarantee (Section~\ref{proof}) through a newly constructed bit-flipping mechanism,  and a provable utility guarantee (Section~\ref{proof_utility}) through an estimation of the recovery error for ACIs. The combination of bit-flipping, DP, and quantization outperforms existing methods across various metrics and provides a leakage-resilient, communication-efficient, and carbon-neutral solution.

\item We evaluate the model utility, carbon emissions, and communication cost of \tool. We found that \tool can halve communication cost with minimal carbon emissions (estimated through computational overhead) and negligible impact on model performance.

\item We comprehensively evaluate \tool against state-of-the-art data reconstruction attacks, such as the ROG attack in CV tasks and the LAMP attack in NLP tasks. The proposed bit-flipping mechanism introduces controllable noise with a larger variance while providing provable differentially private protection for ACIs. \tool demonstrates superior defense performance compared to benchmarks. We also demonstrate that solely applying gradient quantization or differential privacy is challenging to achieve a satisfying trade-off between utility and defense performance.
\end{itemize}

To the best of our knowledge, CDPA is the \textit{first} work that introduces a practical framework for ACIs that provides privacy and efficiency guarantees simultaneously, particularly focused on two trade-offs: the trade-off among accuracy, carbon emissions, and communication overhead, and the trade-off between privacy and utility. This enables its applicability to new and emerging domains such as space, 6G networks, and agriculture, aligning with contemporary digital technological advancements.

\section{Related Work}
\label{background}

In this section, we introduce the background knowledge and existing research related to the context of ACI to provide protection against data leakage and/or alleviate communication bottlenecks. We compare several related studies in Table~\ref{tab_related_work}. More details can also be found in Section~\ref{sec_attack_and_benchmark} and~\ref{sec_attack_and_benchmark2}. 

\subsection{Data Leakage Risks in ACIs}

The safety, privacy and resilience of critical infrastructure is a national security issue~\cite{vigano2020cybersecurity}. Critical infrastructure systems often collect vast amounts of data, including personal information, and whose failure could cause catastrophic loss of life, assets, or privacy~\cite{laplante2021artificial}. For example, AI-powered surveillance systems can monitor and track individuals in critical infrastructure environments, potentially infringing on their privacy rights~\cite{fontes2022ai}. 
Several studies look into threats to the privacy and security of critical infrastructures, such as jamming attack~\cite{harrison2018space,ohlmeyer2006analysis,rausch2006jamming}, spoofing~\cite{pavur2020tale,schmidt2016survey}, and distributed denial-of-service~\cite{giuliari2021icarus,usman2020mitigating}. 

Recent study reports that even when an attacker is passive with regards to satellite signals and cannot directly inject, spoof, or interrupt radio emissions, he/she can reliably interpret a significant portion of broadcast data from satellite communication~\cite{pavur2020tale}. Therefore, to apply FL in critical infrastructure, a major concern is the potential exposure of private information via parameter uploads, such as membership inference attacks~\cite{carlini2022membership,shokri2017membership,salem2018ml,chen2020gan,nasr2019comprehensive}, attribute inference attacks~\cite{melis2019exploiting,song2019overlearning}, and data reconstruction attacks~\cite{zhu2019deep,zhao2020idlg1,geiping2020inverting,yin2021see,yue2023gradient,lu2022april,hatamizadeh2022gradvit,li2022auditing,deng2021tag,balunovic2022lamp}. 
In this study, we particularly focus on proposing an effective privacy protection against data reconstruction attacks since they attempt to reconstruct or infer sensitive information from a trained machine learning model, revealing private or confidential information about individuals in the dataset. 
To address the data leakage issue, local differential privacy (LDP) has been applied to protect private information during the neural network training process~\cite{abadi2016deep,wei2020federated,miao2022compressed} (as detailed in Appendix ~\ref{app_dp}). %
Soteria~\cite{sun2021soteria} maximizes the distance between the original and reconstructed images by pruning the single layer gradient and perturbing the data representation in the model. We show that relying solely on differential privacy presents challenges in achieving a satisfactory balance between utility and defense performance.

\begin{figure*}[t]
\centering
\includegraphics[width=0.9\linewidth]{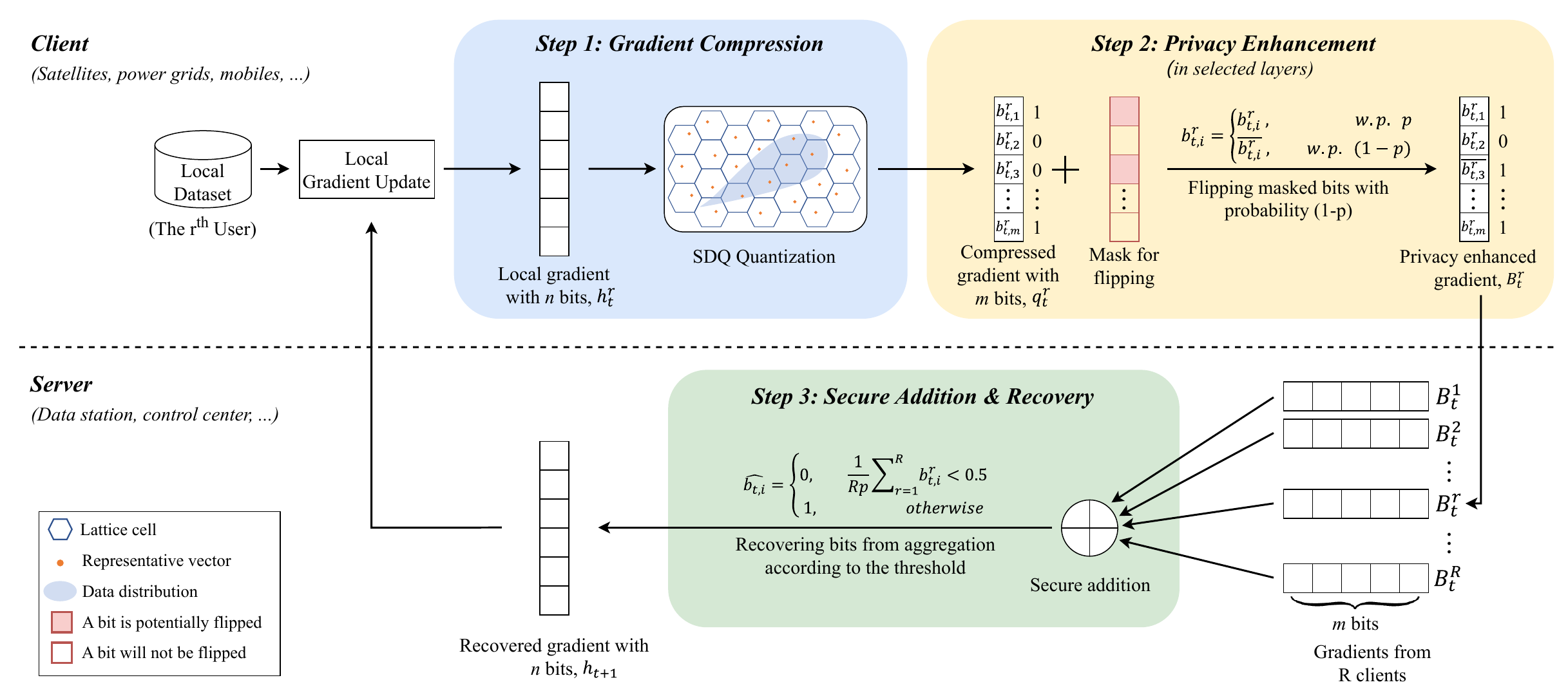}
\caption{Overview of \tool. During the client-side gradient quantization process, gradient updates are quantized at a fixed rate. Then, in selected layers, each bit in the binary representation is flipped according to a flipping mask with probability $1-p$. 
At the server-side, binary data are aggregated and then recovered. }
\label{fig_overview}
\end{figure*}

\subsection{Communication Bottlenecks in ACIs}

In ACIs, ensuring robust and efficient communication throughput is paramount. This involves optimizing data transfer and exchange, leveraging artificial intelligence to enhance reliability, reduce latency, and bolster overall performance. However, recent research reveals that the complex interactions of distributed applications and the low speed of communication lines pose obstacles to the efficiency of management and timely decision-making within critical infrastructure~\cite{kosenko2017methods,kodheli2020satellite}. 
For instance, the application of high-resolution sensors results in  conventional approach of transmitting captured images to the ground for analysis and distribution no longer practical due to the vast amounts of data generated by Earth observation satellites, prompting the need for novel processing techniques~\cite{bui2023board}. Several studies focus on either improving the processing capabilities~\cite{tang2021computation}, optimizing the communication and computation resources to reduce the latency~\cite{gost2022edge,cui2020latency}, or extending network architectures~\cite{wang2018computation}.

On the other hand, as discussed previously, FL has achieved great success in training
models in a distributed manner. However, this may not necessarily provide a perfect solution to the problem, as even the transmission of gradients could put too much pressure on the capacity of critical infrastructure networks, especially considering the rapid increase in the scale of machine learning models. 
Recent studies utilize gradient compression or quantization technology~\cite{bernstein2018signsgd,bernstein2018signsgd2,aji2017sparse,alistarh2017qsgd,shlezinger2020uveqfed,lang2023joint} to address this issue.
GradDrop~\cite{aji2017sparse} reduces communication cost by sparsifying the gradients, however, a high compression loss exists. 
QSGD~\cite{alistarh2017qsgd} provides unbiased gradient estimations at the scalar level, while signSGD~\cite{bernstein2018signsgd,bernstein2018signsgd2} offers biased estimations with only sign signals preserved. However, both approaches can introduce considerable compression or quantization errors, affecting model convergence and performance. Given this, we incorporated SDQ~\cite{shlezinger2020uveqfed}, which has demonstrated superiority in minimizing quantization errors, as an integral part of our methodology to enhance communication efficiency.

\section{Threat Model}
\label{sec_threat_model}
In this section, we define our threat model according to the methodology by Carlini~\etal~\cite{carlini2019evaluating} and describe this threat model with the goals, capabilities, and knowledge of the adversary.

\noindent \textbf{Adversary's goal.~} 
In our threat model, the adversary attempts to obtain the client's private data $\mathcal{D}$ without accessing them directly. Specifically, the adversary may reconstruct training examples that are close to raw ones. As discussed previously, such private data in ACIs could include operational data, user information, sensitive records, or even security credentials. In this study, we specifically consider image and natural language datasets as examples of private data. 

\noindent \textbf{Adversary's capability.~} Following related studies~\cite{yue2023gradient,balunovic2022lamp}, we consider an honest-but-curious server that is compromised by the adversary. Therefore, the attacker is capable of compromising a server in ACI networks. 
The server, even being compromised, is not allowed to modify the ACI model training protocol \eg in each federated learning iteration $r$, it will collect gradients $h_t^r$ from clients, aggregate and train the global model $M$, and send back updated gradients $h_{t+1}$ to clients honestly. We did not consider data poisoning attacks that compromise clients, since several recent studies~\cite{jia2021intrinsic,jia2022certified,levine2020deep,jia2023pore} provide defense strategies that can be applied to our framework (as detailed in Section~\ref{potential}). 

\noindent \textbf{Adversary's knowledge.~}
We assume that the adversary has white-box access to the server through compromising. Given that FL employs a training approach where all clients share a common global model, it becomes straightforward for an adversary to discern the model architecture $M$ and gradients communicated between clients and server, \ie $h_t^r$ and $h_{t+1}$. To further challenge our defense mechanisms, we assume that the adversary has full knowledge to the ground-truth label set. 
Therefore, by iterating over all possible labels, the adversary can achieve the best reconstruction outcome.

\section{\tool Methodology}
In this section, we introduce the methodology of our proposed \tool, which aims to provide a defense scheme that enhances model privacy while simultaneously improving model utility and communication \& computation efficiency.

To address the challenge of communication bandwidth and cost in ACIs, we integrate vector lattice quantization into \tool, which efficiently compresses and encodes the parameters (\eg gradients) for transmission. Such design reduces the volume of data exchanged between clients and the server, allowing for streamlined communication in ACIs while minimizing the associated expenses.
Our privacy strategy achieves differential privacy through a novel bit-flipping approach. More specifically, we have designed a random bit-flipping mechanism, which is applied to the binary representations of parameters with a certain probability, rather than directly adding noise to all parameters. This mechanism offers superior privacy protection while incurring lower computational costs.
As shown in Figure~\ref{fig_overview}, there are three main steps designed in \tool, including Gradient Quantization, Privacy Enhancement, and Secure Addition \& Recovery.
\subsection{Step 0: Initialization}
 
The initialization of the model follows the conventional approach in federated learning, where all initial models are deployed on clients (\eg satellites, power substations, marine cargo, and agricultural monitors) to share the same set of weight parameters during the first training iteration. The server (\eg ground station, energy grids control center, and maritime operation center) will collect gradient updates, rather than raw data, from each client. In each epoch, the server will aggregate the gradients and train a global model, and then send the updated global gradients back to clients. More details are provided in Appendix ~\ref{app_fl}. 
\subsection{Step 1: Gradient Quantization}\label{gradient_quantization}

Compared to other distributed machine learning approaches, FL shows promise in significantly reducing communication costs in ACI networks because it does not require the transmission of raw images or samples. However, even transmitting gradients alone can strain ACI networks, especially with larger machine learning models. Therefore, we consider gradient quantization as a potential solution. Specifically, we incorporated Subtractive Dithered Lattice Quantization (SDQ)~\cite{shlezinger2020uveqfed,shlezinger20221} (as detailed in Appendix ~\ref{sdq_background}) 
into the \tool framework, as it introduces fewer quantization errors compared to other quantization methods. \textbf{We note that our framework is flexible, and other quantization approaches could also be seamlessly integrated}, but they may contribute differently to the final reconstruction error (refer to Section~\ref{sec_secure_addition_and_recovery} for more details). As demenstrated in Algorithm~\ref{alg_CDPA}, for each client $r$, we first quantize its local gradients $h_t^r$ in each epoch $t$ with SDQ to obtain the quantized gradient $q_t^r$ (line~\ref{line_sdq}). After quantization, $h_t^r$ with $n$ bits will be compressed into $q_t^r$ with $m$ bits (\ie quantization rate $a = n/m$). 

\begin{algorithm}[t]\footnotesize
\caption{\tool in selected layers.}
\label{alg_CDPA}
\KwIn{A set of clients $R$, local gradient updates from $r^{th}$ client in epoch $t$ iteration $h_t^r$, quantization rate $a$, bit-flipping mask $s$, flipping probability $1-p$, lattice $\mathcal{L}$, bit recover threshold $T$.}
\KwOut{Global model gradient update $h_{t+1}$ in epoch $t+1$.}
\tcc{At client-side}
\For{$r \gets 1$ to $R$}{
    $q_t^r \gets \sdq(h_t^r, a, \mathcal{L})$\;\label{line_sdq}
    $B_t^r \gets \floatToBin(q_t^r)$\;\label{line_float2bin}
\For{each $b_{t}^r \in B_t^r$}{\label{line_for_bt}
    \If{$\isMasked(b_{t}^r, s)$}{
        $b_{t}^r \gets \bitFlip(b_{t}^r, p)$\;\label{line_bit_flip}
    }
}
$\sendToServer(B_t^r)$\;\label{line_send2seerver}
}
\tcc{At server-side}
$\widehat{B}_t \gets \phi$\;
\For{$r \gets 1$ to $R$}{
    $\widehat{B}_t \gets \secureAddition(\widehat{B}_t, B_t^r)$\;\label{line_secure_addition}
}
\For{each $\widehat{b_t} \in \widehat{B}_t$}{\label{line_for_bt_hat}
    \If{$\recover(\widehat{b_{t}}) < T$}{\label{line_recover}
        $\widehat{b_{t}} = 0$\;
    }
\Else{$\widehat{b_{t}} = 1$\;} \label{line_recover2}
}
$h_{t+1} \gets \binToFloat(B_t)$\;\label{line_bin2float}
\end{algorithm}

\begin{figure}[t]
\centering
\includegraphics[width=\linewidth]{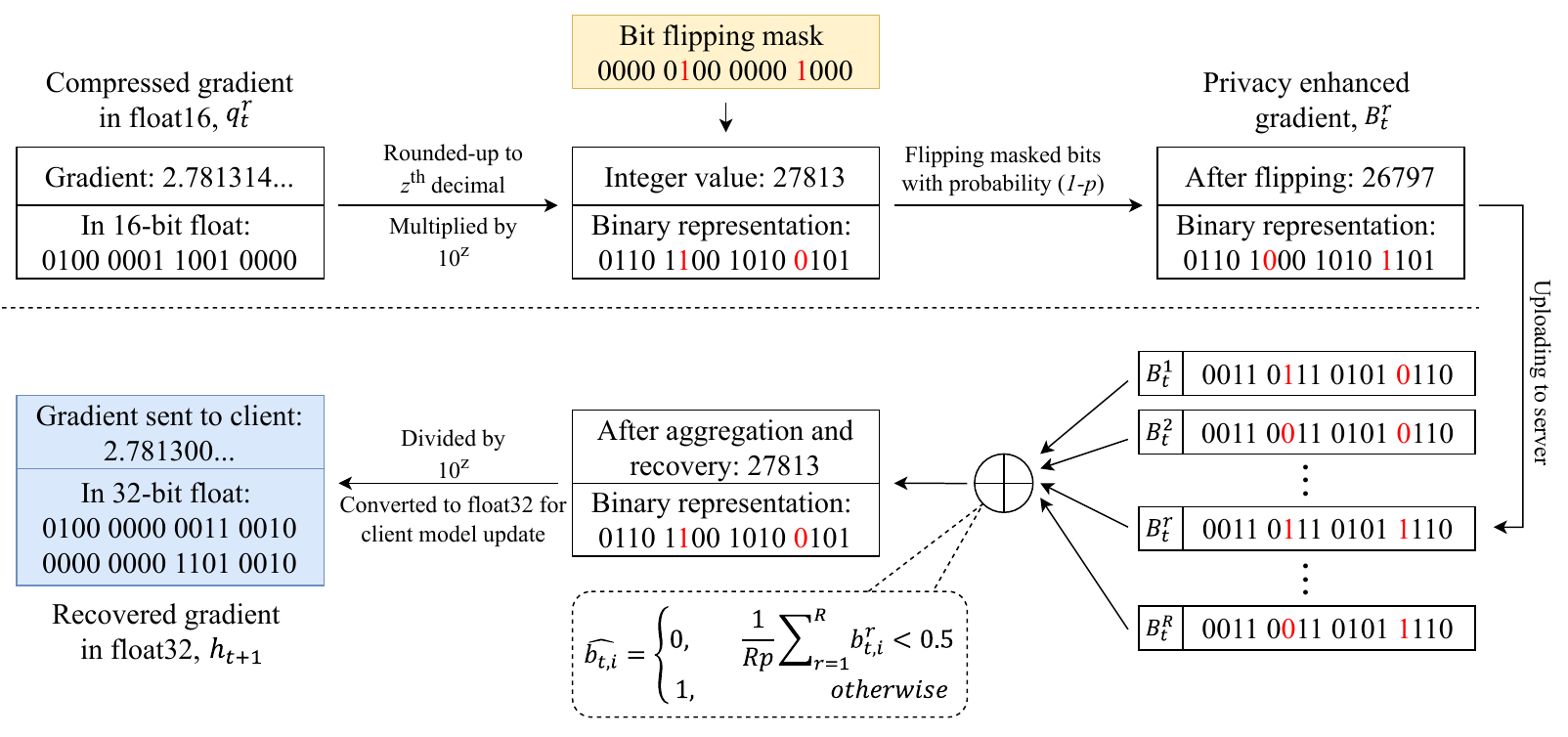}
\caption{An example of gradient encoding (bit-flipping) on the client-side and decoding (aggravation with flipping restoration) on the server-side.
While the bit-flipping may affect the output of a particular client, the restoration process aims to restore the gradient. Specifically, the flipping restoration aggregates and restores the masked bits that have been flipped (highlighted in red) according to the flipping probability $1-p$. The accuracy of the decoding process increases with the number of clients participating in the process, denoted by $R$.}
\label{fig_bit_flipping}
\end{figure}

\subsection{Step 2: Privacy Enhancement}

The privacy enhancement process is carried out on the client-side after the quantized gradient $q_t^r$ is obtained. We would like to note that gradient quantization could also (to a limited extent) contribute to privacy enhancement, as it adds perturbations to the transmitted data.
To propose a strong defense against potential privacy and security attacks on ACI networks (\eg data reconstruction attacks), with the challenge of communication cost in mind, we have particularly designed a random bit-flipping mechanism to perturb the gradients transmitted from clients to the server, thus protecting them from being leaked. 

\noindent \textbf{Bit-flipping.~} 
We see our bit-flipping mechanism as a special DP technique, which adds noise to the quantized gradient $q_t^r$ through flipping specific bits with a probability $1-p$ according to a bit flipping mask $s$ (lines~\ref{line_for_bt} to~\ref{line_bit_flip} in Algorithm~\ref{alg_CDPA}). If the bit at the $i$-th position (starting from $0$ on the left side) in a binary presentation is masked (\ie set as \textcolor{red}{1} in $s$). 
The $\bitFlip()$ in line~\ref{line_bit_flip} is designed as follows:
\begin{equation}
 {b}_{t,i}^r=\left\{\begin{array}{cc}
 {b}_{t,i}^r, & w.p.~p \\
\overline{{b}_{t,i}^r}, & w.p.~(1-p),
\end{array}\right.
\label{equ_bit_flip}
\end{equation} 
where ${b}_{t,i}^r$ is the bit at the $i$-th position in a binary representation (\ie a gradient from epoch $t$), and $w.p.~(1-p)$ means flipping the bit with probability $1-p$.

The bit-flipping adds noise to the gradients, which protects them from being leaked during data transmission from the client to the server. However, it is critical to determine the number of bits to be flipped, as flipping too many bits may introduce too much noise and significantly affect the training efficiency of the model; while flipping too few bits may lead to less noise and limits the defense performance. Note that since each bit in binary represents different values according to its position, flipping bits on the left will introduce noise with a larger variance. Therefore, unlike traditional DP, the noise added by the proposed bit-flipping mechanism follows a discrete distribution, \ie flipping a bit may introduce a constant noise. This enables the gradient recovery during aggregation on the server-side, as introduced in Section~\ref{sec_secure_addition_and_recovery}.
Specifically, the bits to be flipped are controlled by the flipping mask $s$, defined as a hyperparameter during the training of the model. In our experiments, we empirically set two bits as \textcolor{red}{1} in the mask. We note that such a decision is flexible, and in practice, the mask should be determined based on the desired noise level to be added according to a specific scenario. 

\noindent \textbf{Binary conversion.~}
To control the noise-level in a convenient manner, in the implementation of bit-flipping, we first convert $q_t^r$ from float type into a specific designed binary representation (line~\ref{line_float2bin} in Algorithm~\ref{alg_CDPA}). According to IEEE 754~\cite{markstein2008new}, a number in 16-bit floating-point format consists of three parts: \texttt{Sign} as the 1st bit, 5 bits for the \texttt{Exponent} part, and 10 bits for the \texttt{Fraction} part. The numerical value of a floating-point number is calculated as $(-1)^{\texttt{Sign}} \times 2^{\texttt{Exponent}} \times (1+\texttt{Fraction})$. Therefore, flipping a bit in a floating-point number may lead to unstable noise, making it difficult to control the noise level. Inverting the \texttt{Fraction} bit might not significantly alter a gradient value, rendering it ineffective against data reconstruction attacks. On the other hand, modifying the \texttt{Exponent} bit could excessively change the conveyed gradient values, potentially hindering model convergence. Hence, a uniform and controllable precision conversion is foundational for the \tool scheme.

To address the aforementioned challenges, we proposed a specifically designed binary conversion approach. We first round all gradient values to the $z$-th decimal place after SDQ, and subsequently multiply the gradient $q_t^r$ by $10^z$. 
As shown in Figure~\ref{fig_bit_flipping}, when taking $z=4$, a gradient of approximately ``$2.781314$'' (represented as ``0100 0001 1001 0000'' in 16-bit float) will be convert to ``0110 1100 1010 0101'' (or ``27813'' in decimal). 
Through such conversion, we can ensure that each bit represents consistent precision. In other words, flipping a bit in the converted binary will result in stable noise of exponential of 2. It is important to note that this conversion is primarily for conveniently controlling the noise level. While it is possible to directly flip bits in a floating-point number, doing so might require a more complex design of the flipping mask, which is not the focus of this study.
Then the amplified gradients is recovered through dividing by $10^z$. 

In the defense scheme, we assume that attackers are aware of the value of $z$ in our novel binary conversion. This implies that they can obtain the amplification factor $z$ used in the encoding process, enabling them to decode the transmitted binary integer back to its corresponding decimal floating-point representation.

\noindent \textbf{Target layer selection.~}
To reduce computational cost, we only apply \tool on selected layers in a machine learning model.
As highlighted in previous research~\cite{sun2021soteria}, in CNN-based models, the last fully connected layer is the most susceptible layer to data reconstruction attacks, since it contributes more to the final prediction, while other layers may primarily focus on local information (such as effective feature learning in convolution layers).
In another case, for transformer-based models, which have been shown to be more vulnerable than their convolution-based counterparts~\cite{zhang2022does}, the primary source of leakage is the stem layer, self-attention layers, and activation layers, rather than limited to the fully connected layer. Hence, bit-flipping is applied across all layers for these models. Given that these layers encompass nearly the entire model, to streamline the defense strategy, we apply the bit-flip operation to all layers of the model.

\subsection{Step 3: Secure Addition \& Recovery}\label{sec_secure_addition_and_recovery}

The Secure Addition and Recovery step is carried out on the server-side as depicted in Figure~\ref{fig_overview}. After the server receives all client updates $\{B_t^1,...,B_t^R\}$ by performing an new aggregation operation on binary representative manner. Compared with FedAvg~\cite{mcmahan2017communication}, the design of our aggregation mechanism is based on binary computation. The proposed aggregation within binary representation not only enhances the efficiency of the overall operation but also enables the reconstruction of ${\widehat{\mathbf{h}_{t+1}}}$ to recover the ``noise'' introduced by the bit-flipping during the encoding process on client-side, formalized as follows:
\begin{equation}
\widehat{b_{t,i}}= \left\{\begin{array}{cc}
0, & \frac{1}{Rp} \displaystyle\sum_{r=1}^{R} b_{t,i}^r < 0.5\\
1, & Otherwise
\end{array}\right.
\label{agg}
\end{equation} 

We demonstrates the specific process of secure addition and recovery in Figure~\ref{fig_bit_flipping} with a simplified example, which aggregates one parameter from multiple clients. 
The bits marked in red are the bits that are going to be flipped with a probability $1-p$. 
Secure addition is the step of adding 1s and 0s in that position across clients, \ie $\sum_{r=1}^{R} b_{t,i}^r$ in Equation~\ref{agg}. Then, the recovery is applied based on the bit-flipping probability~$1-p$, where the aggregated value is averaged by the number of clients $R$ and the probability $p$. This yields the specific value of $\widehat{b_{t,i}}$, and finally provides an estimation of the binary representation of the gradient updates for a particular position in the aggregated global model. 

\noindent \textbf{Capability of error recovery.~}
Next, we demonstrate that our Secure Addition \& Recovery step can theoretically reduce the error caused by bit-flipping to 0, thus recover the gradients with minimum errors. 
Through the whole process of \tool, there are 3 types of errors involved (neglecting the loss during the communication process), \ie \one~the quantization error $\gamma_1$ introduced by the gradient compression (line~\ref{line_sdq} in Algorithm~\ref{alg_CDPA}), \two~the bit-flipping error $\gamma_2$ caused by the privacy enhancement (lines~\ref{line_float2bin} to \ref{line_bit_flip} in Algorithm~\ref{alg_CDPA}), and \three~the aggregation error $\gamma_3$ caused by the addition and recovery (lines~\ref{line_secure_addition} to \ref{line_recover2} Algorithm~\ref{alg_CDPA}). 

To ensure the utility of \tool{}, we aim to minimize the overall error, denoted as $\gamma=\gamma_1+\gamma_2+\gamma_3$. This is equivalent to minimizing $\gamma_1$ and $(\gamma_2+\gamma_3)$, as $\gamma_2$ can be recovered by $\gamma_3$, but $\gamma_1$ is solely determined by a specific quantization technique.
On another hand, to ensure the privacy of \tool{}, we require that $\gamma_2$ provides guaranteed privacy. Thus, we address utility-preservation and privacy protection simultaneously by \one~selecting a quantization technique with the smallest possible quantization error $\gamma_1$ (as detailed in Section~\ref{gradient_quantization}), \two~designing bit-flipping with guaranteed Differential Privacy (as proven in later Section~\ref{proof}), and \three~ensuring that an appropriate selection of $p$ can result in a sufficiently small $(\gamma_2+\gamma_3)$. 
To simplify the problem, we consider $( \gamma_2 + \gamma_3 )$ for the aggregation of one parameter involving $R$ clients, where only a single bit position is flipped. Then, the probability that the value after recovery is consistent with FedAvg can be presented as 
\begin{equation}\footnotesize
p_{\gamma_2} = \sum_{i = \lceil R/2 \rceil}^{R}\left[\binom{R}{i}p^{i} \times (1-p)^{R-i} \right].
\label{6}
\end{equation}
Hence, the value of $(\gamma_2+\gamma_3)$ is approximately given:
\begin{equation}
 \gamma_2 + \gamma_3 \approx \xi \left(1- p_{\gamma_2} \right) ,
\label{7}
\end{equation} 
where  $\xi = \frac{2^{j-1}}{10^z} $ is a scaling factor depending on the flipped bit position $j$. 

\begin{figure}[t]
\centering
\includegraphics[width=0.8\linewidth]{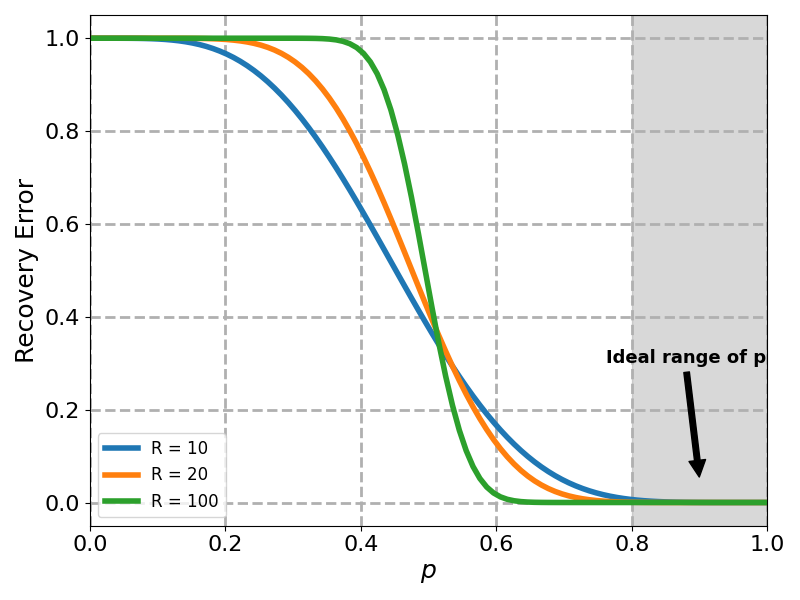}
\caption{Relationship between recovery error induced by bit-flipping and probability $p$.}
\label{fig_p}
\end{figure}

From Equation~\ref{7}, we can ensure that there always be a $p$ that can recover the error $(\gamma_2+\gamma_3)$ to a small value near 0.
We further plot the value of recover error $(1-p_{\gamma_2})$ with different number of clients $R$ in Figure~\ref{fig_p}.
A higher $R$ can further relax the selection boundary of $p$ to obtain a small $(\gamma_2+\gamma3)$. 
Therefore, it is guaranteed that the proposed approach can recover the bit-flipping with an appropriate selection of $p$. More experiments regarding to the selection of $p$ are present in Appendix ~\ref{p_selection}. 
However, we would like to note that, in practice, the selection of $p$ depends on specific trade-off between model utility and privacy protection, that is, a higher utility can be obtained through selecting a larger $p$ that leads to less flipping (\ie less $\gamma_2$ error); while a more robust privacy protection can be achieved by a smaller $p$ which introduces a higher probability of flipping and a larger $\gamma_2$ error.
 
\subsection{Privacy Guarantee of Bit-Flipping} \label{proof}
We adopt differential privacy, a rigorous notion of privacy introduced by Dwork~\etal~\cite{dwork2016calibrating} and widely adopted. It ensures that the algorithm's output is independent of any individual's specific data. This definition enables clients to assess the increased risk associated with participating in a service and make informed decisions about sharing their data.
Formally, a randomized algorithm $M: B^n \rightarrow Y$ satisfies $\epsilon$-differential privacy~\cite{dwork2016calibrating} if, for all neighbouring datasets (\ie clients' values) denoted by $B \sim B'$, we have
\begin{equation}
Pr[M(B) \subseteq T] \leq e^\epsilon Pr[M(B') \subseteq T],
\end{equation}
where $Y \subseteq T$ denotes all possible output, and
$B, B' \subseteq B^n$ differ in exactly one entry at $B_j$.
We will show that the bit-string $M(B_1, ... , B_n) = (Y_1, ... , Y_n)$ is differentially private when our algorithm $M$ is randomized response (RR). According to Equation~\ref{equ_bit_flip}, with the proposed bit-flipping randomized response, a user who possesses a private bit, transmits it correctly with probability $p > 1/2$ or transmits it incorrectly with probability $1-p$. We consider any particular realization $a \in \{0, 1\}^n$ of $(Y_1, ... , Y_n)$. We have that $Pr[M(B) = a] = \prod_{i=1}^nPr[Y_i = a_i]$. Suppose that $B$ and $B'$ differ only in coordinate $j$. 
Let $RR$ be the ratio of two conditional probabilities, \ie $RR = \frac{Pr[M(B) = a]}{Pr[M(B') = a]}$. To satisfy the differential privacy condition, this ratio must be bounded by $e^\epsilon$.
Then we have that
\begin{equation}\footnotesize
\begin{split}
RR & = \frac{Pr[M(B) = a]}{Pr[M(B') = a]} \\
& = \frac{\prod_{i=1}^nPr[Y_i = a_i]}{\prod_{i=1}^nPr[Y_i' = a_i]} \\
& = \frac{Pr[Y_j = a_j]}{Pr[Y_j' = a_j]} \\
& \leq \frac{p}{1-p},
\end{split}
\end{equation}
and
\begin{equation}\footnotesize
\epsilon = \ln(\frac{p}{1-p}).
\end{equation}
Therefore, our bit-flipping RR satisfies $\ln(\frac{p}{1-p})$-LDP.
We note that, the RR here can be viewed as a noise-adding mechanism to obfuscate individual data while enabling the computation of aggregate statistics.

\subsection{Utility Guarantee of \tool}
\label{proof_utility}
In this part, we provide a theoretical proof of \tool utility guarantee, compared to other DP approaches. We utilize assumptions from prior studies ~\cite{thudi2022necessity,hu2019fdml,warnecke2021machine}, focusing on Lipschitz continuity and convexity, which are crucial for trainable deep neural networks like ResNets. There is a main observation for trainable deep neural networks that need flat minima and broad convex regions for effective training, avoiding the difficulties posed by chaotic, non-convex loss surfaces~\cite{li2018visualizing}. For further details on Lipschitz continuity, see Chaudhuri~\etal~\cite{chaudhuri2011differentially}. Additionally, these assumptions are recognized in optimization literature ~\cite{andriushchenko2022towards,wang2024dpadapter} and are essential for trainable neural networks with smooth activation and losses.

\noindent \textbf{The setup for \tool theoretical framework.~}
We briefly simplify the proof by making a few assumptions:
\begin{itemize}[leftmargin=*]
\item \textit{Assumption 1~} The function $f$ of model is $\beta$-smooth with respect to $ \| \nabla f(x|\bm{\theta} ) \|_2 - \nabla f(x|\bm{\theta'}) \|_2 \leq \beta \| \bm\theta - \bm{\theta'} \|_2$.
\item \textit{Assumption 2~} The Loss function $L$ of model is $\beta_1$-Lipschitz continuous with respect to $ |L(x,y) - L(x',y)| \leq \beta_1 |x - x'|. $
\item \textit{Assumption 3~} The function $\ell$ is $\beta_2$-smooth with respect to $x$, \ie $
\|\nabla_x L(x',y)\| \leq \beta_2 |x - x'|.$
\item \textit{Assumption 4~} Considering the gradient vector $\nabla L_i(\bm\theta)$, we assume that the gradient is uniformly $\sigma^2$-subgaussian across the observations $i=1,\ldots,n$, in the sense that for all $i$, it holds that $\mathbb{E}\left[\left\|\nabla L_i(\bm\theta) - \mathbb{E}[\nabla L_i(\bm\theta)]\right\|^2\right] \leq \sigma^2$.
\item \textit{Assumption 5~}  The function $L_D$ complies with the $\mu$-Polyak-Łojasiewicz inequality, namely, for all parameters $\bm\theta$, it satisfies $\|\nabla L_D(\bm\theta)\|^2 \geq \frac{1}{\mu} \left( L_D(\theta) - \inf_{\bm\theta} L_D(\bm\theta) \right)$.
\end{itemize}

Then, our algorithm updates the aggregated parameter $\bm \theta$, from iteration 1 to $t$. On the server side, we have
\begin{align}\footnotesize
\bm{\theta_{t+1}} &= \bm{\theta_t} - \frac{\sum_{r=1}^{R} \eta \nabla L(\bm{\theta_t} + \nabla_t(\gamma))}{R}, \nonumber \\
\nabla_t(\gamma) &:= \gamma \nabla L(\bm{\theta_t})
\label{002}
\end{align}
where $\eta$ is the learning rate at client side and $\gamma$ is the overall error caused by \tool. Based on Equations~\ref{6} and~\ref{7}, $\gamma$ is a constant with given $R$ and $p$. We further set $\rho = \frac{max|f(x|\theta + \nabla_t(\gamma))- f(x|\theta)|}{\| \Delta \|_2}$, which could simplify to be a constant.
We argue that \tool can be analogue to SAM~\cite{foret2020sharpness} for improving model's generality.
Specifically, the error $\gamma_1 + \gamma_2$ of the parameters can be considered to cause the worst-case perturbation on the updated parameters by $\nabla_t(\gamma) $ for each iteration. The final recovery error $\gamma_3$ can reverse the error induced by $\gamma_2$ to close to zero. Contrary to SAM, \tool does not calculate the exact worst-case perturbation, since the recovery process is entirely handled by the bit-flipping mechanism. This further demonstrates the efficiency of our method.

\noindent \textbf{Convergence of \tool.} \tool leads to convergence guarantees when applying the standard training loss.
\begin{theorem}\label{thm:1}
With parameter selection $\hat{\beta} := (\rho^2 \beta_2 + \beta \beta_1)$, then the \tool algorithm enjoys the following utility bound
\begin{equation}\footnotesize
\begin{aligned}
\frac{1}{T} \sum_{t=1}^{T} \mathbb{E}[L_D(\bm{\theta_t})] - \inf_{\bm{\hat{\theta}}} L_D(\bm{\hat{\theta}}) 
&\leq \mu \cdot \left( \frac{4\left( L(\bm{\theta_0}) - \mathbb{E} L(\bm{\theta_T}) \right)}{T \eta} \right. \\
&\quad \left. + 4T {\sigma}^2 \hat{\beta}^2(\eta + \gamma^2) \right),
\end{aligned}
\label{eq:01}
\end{equation}
\end{theorem}
The full detailed proof of Theorem \ref{thm:1} is available at Appendix ~\ref{apx:thm1}.
It suggusts that the better utility achieves while using larger $p$ value (as close to 1), which is because $\gamma \approx \gamma_1 + \xi(1-p_{\gamma_2})$.

\noindent \textbf{Utility Superiority of \tool.} \tool leads to utility improvement compared to other DP approaches without recovery mechanism. 
\begin{theorem}\label{thm:2}
With parameter selection $\hat{\beta} := (\rho^2 \beta_2 + \beta \beta_1)$, then the \tool algorithm enjoys the following bound of loss and grdient:
\begin{equation}
\begin{gathered}
|L(\bm \theta) - L(\bm {\theta'})| \leq \beta_1 \rho |\bm{\theta} - \bm{\theta'}|  = \beta_1 \rho \xi |1- p_{\gamma_2} |, \\
\| \nabla L(\bm{\theta}) - \nabla L(\bm{\theta'}) \|_2 \leq \rho  \beta_2 \| \bm{\theta} - \bm{\theta'} \|_2 + \beta_1 \| \bm{\theta} - \bm{\theta'} \|_2 \\
=   (\rho  \beta_2 + \beta_1) \| \xi \left(1- p_{\gamma_2} \right)\|_2 ,
\end{gathered}
\label{eq:01}
\end{equation}
\end{theorem}
The proof of Theorem \ref{thm:2}~is provided in Appendix ~\ref{apx:thm2}.
Theorem~\ref{thm:2} suggests that selecting a smaller change in parameters makes the algorithm perform better.  $|\bm{\theta} - \bm{\theta'}|$ of \tool is equivalent to Equation \ref{7}, which is nearly approaching to $0$ when $p$ is selected in $[0.8,1]$. Since there is no recovery mechanism, the error of other DP approaches is equivalent to $1/\epsilon$. This guarantees our utility improvement of \tool. 

We further estimate the generalization error using absolute difference between train loss and test loss under settings of Figure \ref{fig_utility_cv} on top-1 accuracy model, as shown in Table~\ref{tab_generalization_error}, we can observe that the generalization error is even lower than that of the vanilla models. This clearly surpasses the upper bounds of traditional vanilla models, which also empirically proves the advantages of \tool on model utility in both accuracy and generalization error.

\input{Tables/tab_generalization_error}

\section{Experiment Setup} \label{exp}

In this section, we introduce the experiment setup, including the machine learning models and datasets involving CV and NLP tasks, the data reconstruction attacks and benchmarks, and the evaluation metrics with respect to model utility, communication overhead, and defense performance. The more detailed experiment setting shows in Appendix ~\ref{app_ex}. 

\subsection{Models}
In computer vision (CV) tasks, the machine learning models are selected based on the number of training parameters, \ie a simple-level model \textbf{SimpleCNN}~\cite{lang2023joint}, a medium-level model \textbf{ConvNet}~\cite{sun2021soteria}, and an advanced-level model \textbf{ResNet18}~\cite{he2016deep} with over 9,000, 730,000, and 11,173,000 trainable parameters, respectively. Except for specific optimizer techniques, all methods utilize the SGD optimizer. In each global epoch, we design 20 clients to participate in the training process. The global epoch is typically set to 95. The dataset is distributed to clients in a fair and independent and identically distributed (i.i.d.) manner.

In NLP tasks, perform experiments on two sizes of BERT models, \ie \textbf{BERT\textsubscript{base}} and \textbf{BERT\textsubscript{distill}}~\cite{wolf2020transformers} with over 110M and 66M trainable parameters, respectively. 
Unlike CV tasks, we set global epoch to 20 and 30 when using dataset CoLA or RottenTomatoes respectively. The number of clients is set to 4. 

\subsection{Attacks}\label{sec_attack_and_benchmark}
In our experiments, we evaluate our defense \tool against five data reconstruction attacks: DLG~\cite{zhu2019deep}, IG~\cite{geiping2020inverting}, and ROG~\cite{yue2023gradient} for CV tasks; and DLG, TAG~\cite{deng2021tag}, and LAMP~\cite{balunovic2022lamp} for NLP tasks, detailed in the Appendix ~\ref{attaks_exp}. 

\noindent \textbf{Considerations on other potential attacks.} \label{potential}
Regarding adaptive attacks, where attackers are fully informed about our \tool method (\eg flipping probability $(1-p)$ and the flipping positions), we argue that their capability to conduct an effective data reconstruction attack is still hindered. This is because even if such critical information is obtained by an attacker, he/she still cannot recover the accurate gradient for each client, which is vital for data reconstruction attacks~\cite{zhu2019deep,geiping2020inverting,yue2023gradient,deng2021tag,balunovic2022lamp}.  
On the other hand, data poisoning attacks, such as label-flipping~\cite{steinhardt2017certified} and gradient-optimization~\cite{munoz2017towards, fang2020local} attacks, where compromised clients tamper with their data to influence the server's global model, could be a potential concern for federated learning, as well as for our threat model.  
However, such threats have been comprehensively studied, and various existing defense mechanisms~\cite{jia2021intrinsic, jia2022certified, levine2020deep, jia2023pore} have been proposed to identify and exclude malicious client updates during aggregation. These approaches are orthogonal to our approach and can be effectively integrated into the \tool framework.

\subsection{Benchmarks} \label{sec_attack_and_benchmark2}

We also included the following state-of-the-art studies as benchmarks in the performance evaluation of \tool, including methods that either utilize gradient compression including GradDrop~\cite{aji2017sparse}, signSGD~\cite{bernstein2018signsgd,bernstein2018signsgd2} and SDQ~\cite{shlezinger2020uveqfed} or introduce perturbations into gradients ranging from Soteria~\cite{sun2021soteria} to LDP~\cite{wei2020federated}, as well as methods that combine both including CAFL~\cite{miao2022compressed} and JoPEQ~\cite{lang2023joint}.
More detials are provided in Appendix~\ref{defense_benchmarks}. 

\subsection{Datasets}
We conducted classification experiments in CV tasks using two widely used image datasets, CIFAR10~\cite{krizhevsky2009learning} and ImageNet~\cite{deng2009imagenet}; and involve two text datasets in NLP tasks, CoLA~\cite{warstadt2019neural} and RottenTomatoes~\cite{pang2005seeing}. The detailed descriptions of datasets are presented in Appendix ~\ref{datasets}.

\subsection{Evaluation Metrics}
\label{Evaluation Metrics}
We evaluate our models' utility with Top-1 accuracy and loss for CV, and Matthews Correlation Coefficient (MCC)~\cite{chicco2020advantages} for NLP. Communication cost is gauged by the average megabits of model updates per transmission epoch. For CV defense, we use PSNR~\cite{castleman1996digital}, SSIM~\cite{wang2004image}, and LPIPS~\cite{zhang2018unreasonable} to measure reconstructed image quality, while for NLP, ROUGE family metrics~\cite{lin2004rouge} are applied. Carbon emissions are estimated per epoch using the Machine Learning Impact Calculator~\cite{lacoste2019quantifying}. Further details are available in the Appendix ~\ref{app_metrics}. 

\section{Results}
In this section, we report the experimental results, comparing the performance of \tool with benchmarks in terms of model utility, defense performance, communication cost, and carbon emissions.

\subsection{Experimental Results in CV Tasks}
\label{sec_results_cv}

In computer vision (CV) tasks, the experimental outcomes is mainly focusing on four parts: model utility, defense performance, communication cost, and carbon emission estimation. To further confirm the defense performance, we also conduct a user study with 100 participants involved.

\begin{figure}[t]
\centering
\includegraphics[width=\linewidth]{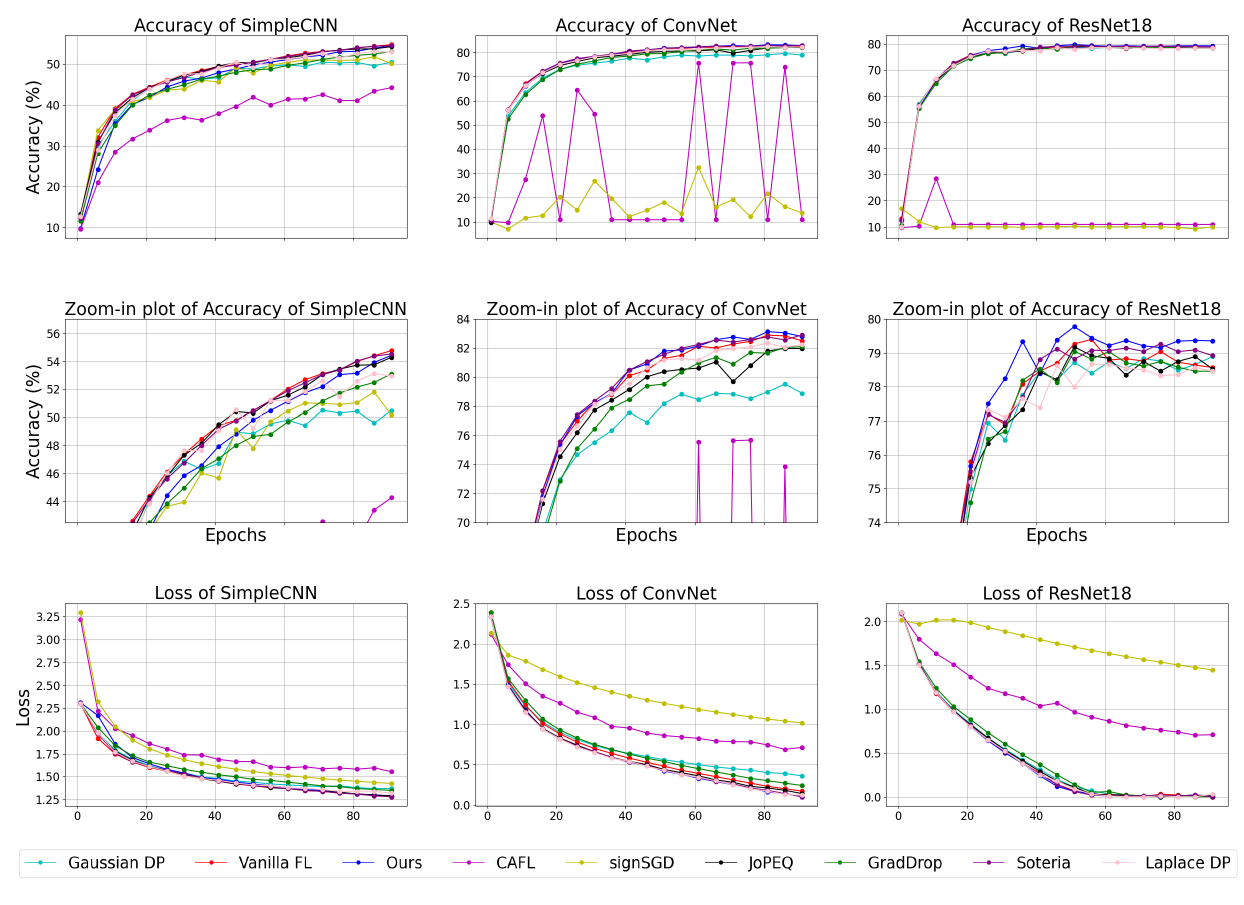}
\caption{A comparison of CDPA with other schemes on model utility. }
\label{fig_utility_cv}
\end{figure}

\noindent \textbf{Model utility.~} 
The model utility result is illustrated in Figure~\ref{fig_utility_cv}. We evaluate the accuracy and prediction loss of \tool and other benchmarks across models with different complexity, \ie SimpleCNN, ConvNet, and ResNet18. The second row in Figure~\ref{fig_utility_cv} presents zoom-in accuracy plots.
Notably, \tool consistently outperforms other strategies in maintaining high accuracy. In the ResNet18 model, \tool has the highest accuracy across epochs. For ConvNet, \tool starts strong and remains competitive in later epochs, while in SimpleCNN, its advantage is particularly pronounced during the latter epochs. From loss perspective, \tool demonstrates commendable stability, maintaining a low loss value in all three models. 

Among the benchmarks, signSGD and CAFL demonstrate limited performance. Particularly when facing a higher model complexity (\eg in ConvNet and ResNet18), these two methods struggle to converge, exhibiting high loss values. In other benchmark methods such as JoPEQ, GradDrop, LDP, and Soteria, we observe a relatively stable performance trend. While these methods can still maintain consistent performance in more complex models, \tool performs the best (54.8\%, 83.3\%, and 79.8\% in SimpleCNN, ConvNet, and ResNet18, respectively).

\input{Tables/tab_defense_ROG}

\begin{figure}[t]
\centering
\includegraphics[width=\linewidth]{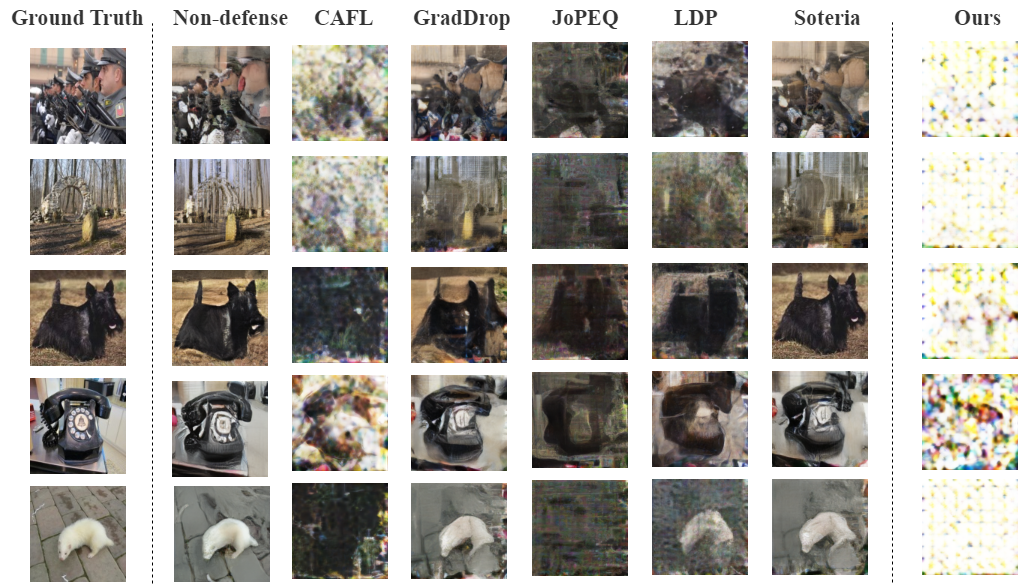}
\caption{Visualization of defense performance against ROG attack.}
\label{fig_Visual}
\end{figure}

\begin{figure}[t]
\centering
\includegraphics[width=0.9\linewidth]{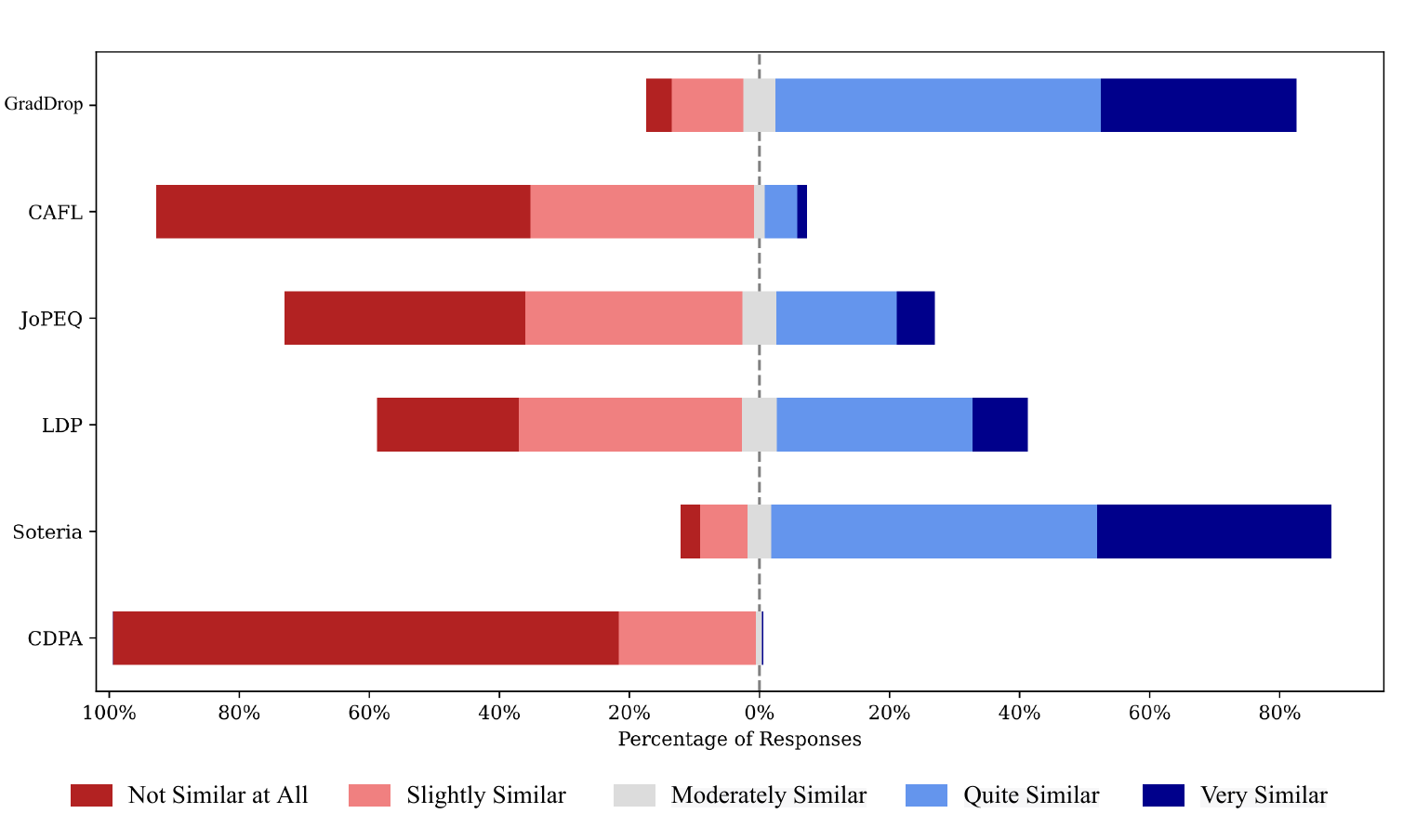}
\caption{User study results on defense schemes against ROG attack. }
\label{user study}
\end{figure}

\noindent \textbf{Defense performance.~} We first evaluate our proposed \tool against ROG attack, using Resnet18 model on ImageNet dataset, and comparing it with other state-of-the-art defense schemes. 
Tables~\ref{tab_defense_ROG} delineates the quantitative performance of various defense mechanisms, utilizing metrics PSNR, SSIM, and LPIPS. Among the benchmarks, \tool recorded the lowest PSNR and SSIM values, registering at 5.114 and 0.190, respectively. These results indicate that the unrecognizability of images reconstructed from models being protected by \tool surpasses those protected by other methods. Specifically, \tool exhibits a significant perceptual difference from the original, as evidenced by a high LPIPS value (0.732), surpassing most competing defenses.
The visualization of reconstructed images shown in Figure~\ref{fig_Visual} further contextualize the experimental results. The reconstructed images from models protected by \tool contain heavily obfuscated image attributes, rendering a superior resilience against data reconstruction attacks. Compared to \tool, other methods such as CAFL, LDP, and JoPEQ may still leak discernible features of the original images, resulting in blurry but still recognizable reconstructions. In contrast, GradDrop and Soteria essentially expose nearly all of the private data.

We further considered the scenario where a stronger attacker has chances of recovering aggregated bits through collecting a large number of clients' gradients (\eg imputes outliers in gradients mean value). However, the randomness inherent in bit-flipping and the one-way recovery process make it challenging for an attacker to precisely reverse a specific client's gradient.  We present experimental results in Table ~\ref{tab_reconstruction} 
in the appendix. 
Even in a malicious-server scenario where the server tends to recover client's weight through de-aggregated weight updates~\cite{pasquini2022eluding,zhao2023loki}, we argue that, \tool, by constantly employing large magnitude of provable DP noise on each client, still sustains a strong defensive performance without compromising.

\noindent \textbf{User study.~}
We further conducted a user study to qualitatively assess the perceptibility of reconstructed images, comparing them to the original images. 
Particularly, we recruited 100 anonymous participants to undertake an online survey via Amazon Mechanical Turk (AMT)~\cite{amazon_turk}. 
In each survey question, we present a pair of reconstructed and original images, and asking how they are similar based on a 5-point Likert scale~\cite{likert1932technique}, \ie ``Not Similar at ALL'', ``Slightly Similar'', ``Moderately Similar'', ``Quite Similar'', and ``Very Similar''. 
For each benchmark method, we randomly selected 10 pairs of images. 
The Human Research Ethics Committee of the authors' affiliation determined that the study did not require additional human subjects review. 
More details of the user study are provided in Appendix ~\ref{app_survey_protocol}. 

As illustrated in Figure~\ref{user study}, when models are protected by \tool, almost 78\% of participants consider the reconstructed images and original images are ``Not Similar at ALL'' and 22\% participants recognized them as ``Slightly Similar'', which indicates that \tool is the most effective method in protecting original images from being reconstructed. 
On the contrary, GradDrop reveals a notable vulnerability as many participants (80\%) found the images to be ``Quite Similar'' or ``Very Similar''. 
Other methods, such as CAFL, JoPEQ, LDP, and Soteria, show varying degrees of similarity feedback, indicating their defense capabilities are unstable on different samples. Yet, for each of them, there is a segment of participants who believe their images are either ``Quite Similar'' or ``Very Similar''. Overall, \tool stands out as the superior choice to ensure data protection.

\begin{figure}[t]
\centering
\includegraphics[width=\linewidth]{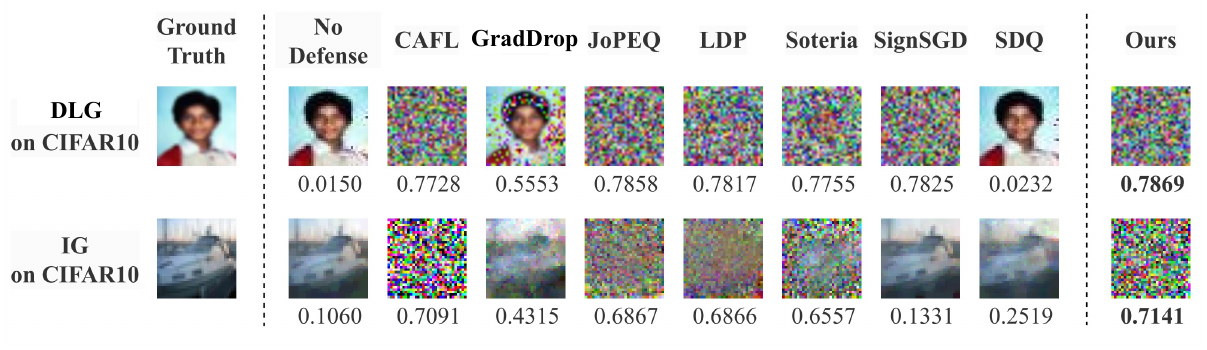}
\caption{Defense performance against DLG and IG attacks. The LPIPS values are shown in the bottom of images.}
\label{idlg}
\end{figure}

We further evaluated the performance of these defensive schemes against DLG and IG attacks. The results are presented in Figure~\ref{idlg}. 
Note that the reported LPIPS for each protection method are average values across all reconstructed samples.
It is shown that different attack methods have varying impacts on the reconstruction quality of the model. The attack capability of the IG attack is generally stronger than that of the DLG attack. 
From the results, we can see that utilizing quantization techniques alone is limited in providing a protection against data reconstruction attacks. Specifically, SDQ, which employs a 16-bit quantization on gradients, results in subpar defensive outcomes (indicated by a low LPIPS 0.2519). 
Even more starkly, signSGD, which uses a 1-bit compression, is highly susceptible to IG attack (LPIPS 0.1331). These observations suggest that \textbf{relying solely on quantization-based approaches might not significantly bolster a model's defensive capabilities}. On another hand, defense mechanisms related to DP exhibit commendable performance against both DLG and IG attacks. Although, in our experiment, GradDrop and Soteria have the same compression rate, Soteria exhibits better defensive results. This can be attributed to the optimization strategies incorporated within the Soteria algorithm. Our proposed method, \tool, is capable of effectively defending against both DLG and IG attacks, demonstrated by the highest LPIPS values (achieving LPIPS 0.7869 against DLG attack and 0.7141 against IG attack). 

\begin{figure}[t]
\centering
\includegraphics[width=\linewidth]{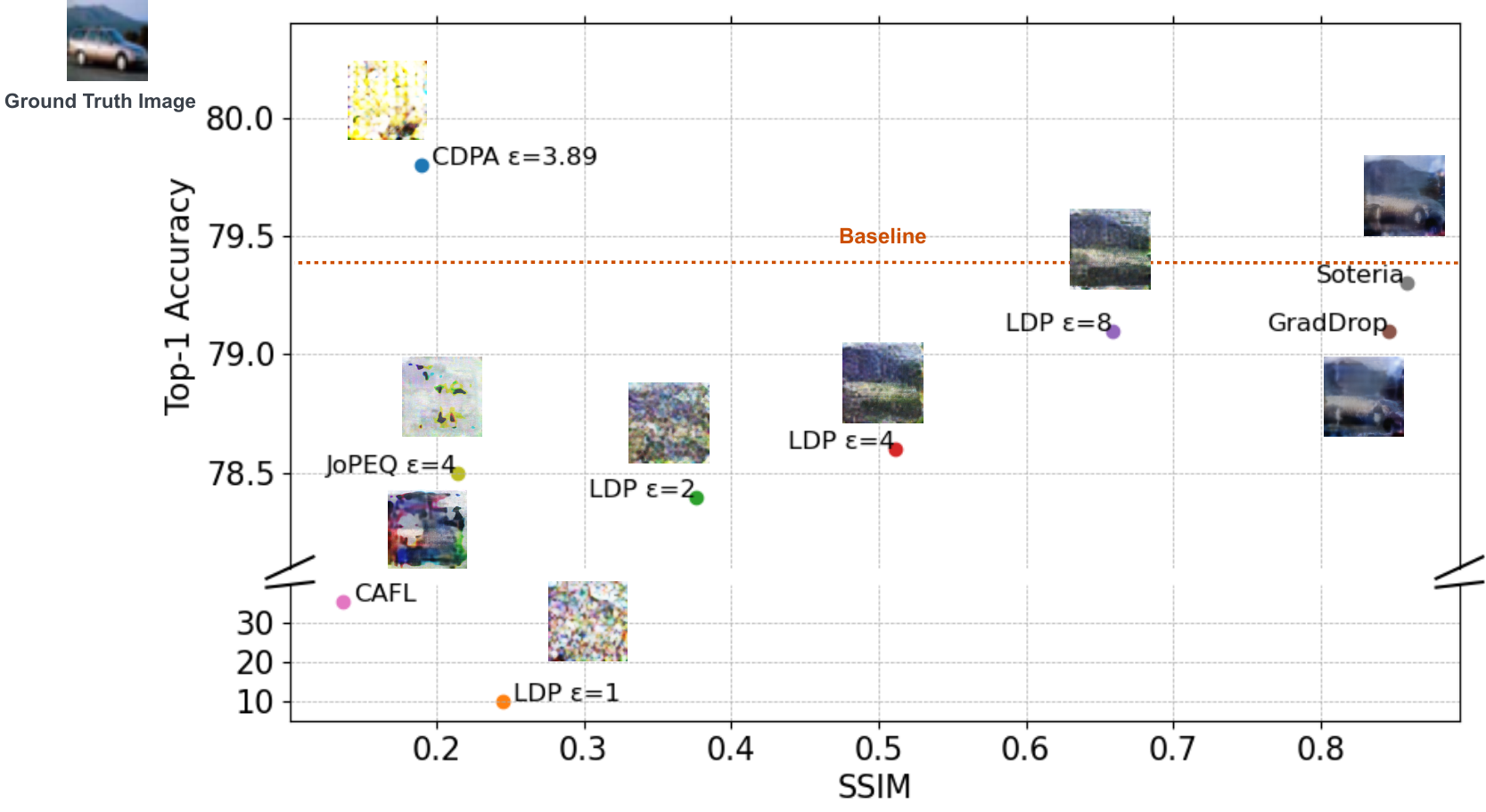}
\caption{Trade-offs between model accuracy and privacy, comparing various defensive methods under the ROG attack. An ideal defensive method should lie in the upper-left corner, demonstrating a superior trade-off between utility and data protection. }
\label{DP_05}
\end{figure}

\noindent \textbf{Model utility \vs defense trade-off.~}
To illustrate the trade-off between model utility and defense performance, we particularly look into ROG attack as a case study. In Figure~\ref{DP_05}, we report the top-1 prediction accuracy and SSIM of various benchmark protection methods.
Ideally, defensive techniques would be situated in the top-left corner, implying high model accuracy coupled with a low SSIM, ensuring robust data protection. Notably, \tool is located in the most upper-left corner, which demonstrated a superior trade-off between utility and privacy protection. In comparison, with gradient compression techniques such as Soteria and GradDrop, we can almost see the fully reconstructed private images. 
For CAFL, although the reconstructed images become more difficult to recognize to some extent, a prediction accuracy near 30\% even affected the model's convergence. We present LDP results with various $\epsilon$ values to demonstrate the defense performance when only differential privacy (DP) is applied. An $\epsilon$ value of 1 results in significantly reduced prediction accuracy. However, with LDP at $\epsilon = 4$ and $8$, some attributes can still be identified from the images, and model accuracy is also lower than that achieved with \tool. This indicates that achieving \textbf{a good trade-off between model utility and defense performance is challenging when applying differential privacy alone in defense} against data reconstruction attacks.
Additional experimental results with varying $p$ and corresponding $\epsilon$ values are provided in Table ~\ref{tab_various_epsilons} 
in the appendix. A higher $p$ may lead to higher accuracy but lower defense performance. However, selecting $p$ within a proper range (\eg $0.8<p<1$ in Figure~\ref{fig_p}) does not significantly affect model accuracy and defense performance.
In conclusion, CDPA demonstrates superior performance with respect to utility and security trade-offs.

\begin{figure}[t]
\centering
\includegraphics[width=\linewidth]{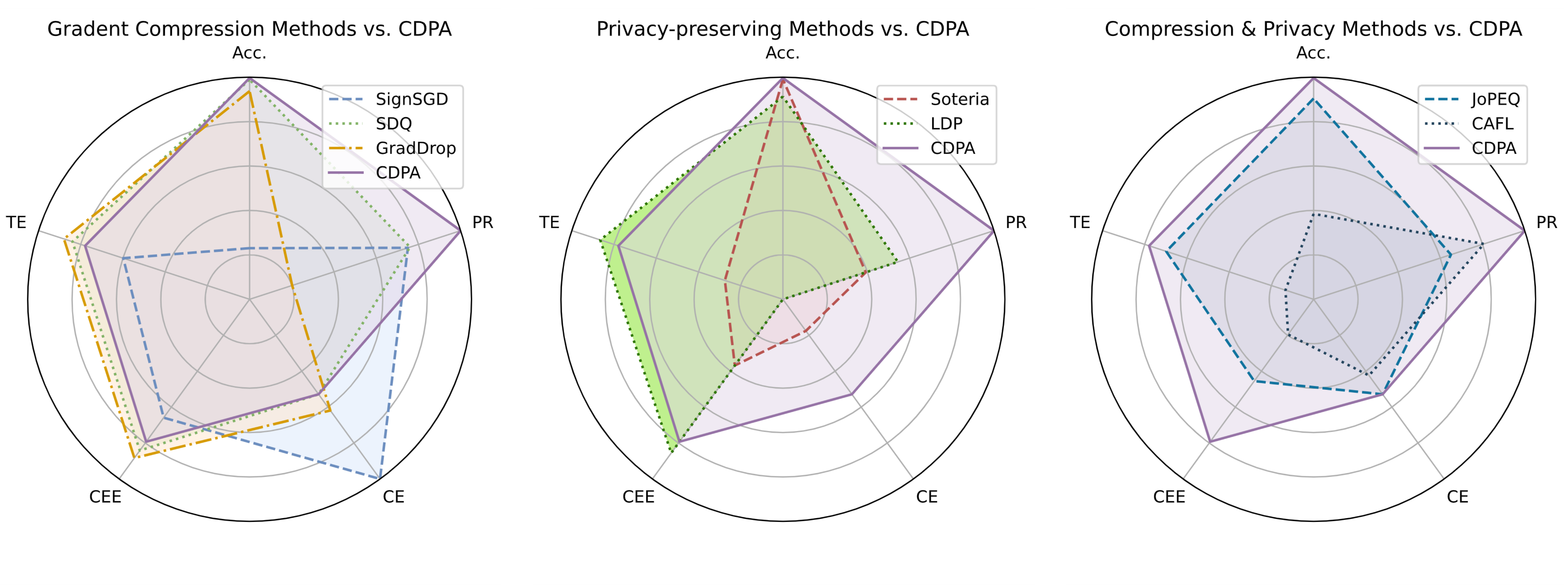}
\caption{Comparison of CDPA with exsiting methods, regarding accuracy (Acc.), training efficiency (TE), carbon emission efficiency (CEE), communication efficiency (CE), and privacy (PR).}
\label{radars}
\end{figure}

\input{Tables/tab_privacy_overhead}
\noindent \textbf{Communication cost.~} Communication efficiency generally refers to the amount of data transmitted from clients to the server per global epoch. We calculate the difference in communication cost between the original model and the models using different methods, as shown in Table~\ref{tab_cv_results}. The \tool scheme manages to reduce the communication cost by half while nearly leaving the model's utility unaffected. Although signSGD can save over 95\% of communication overhead, its model accuracy rapidly declines as the model complexity increases. Soteria achieved a compression ratio of 60\%, but its actual reduction in communication cost is less than 1\%. This is because Soteria only prunes the fully connected layers. GradDrop achieves up to 60\% savings in communication cost, but in Table~\ref{tab_defense_ROG} and Figure~\ref{idlg}, it exhibits fewer capabilities against data reconstruction attacks. Hence, with respect to the communication efficiency of the model, our approach, CDPA, exhibits notable superiority.

\noindent \textbf{Carbon emission estimation.~} The energy cost of privacy refers to the additional computational cost incurred by privacy techniques during the training process. We calculate the difference in carbon emissions between training a vanilla model and different methods to quantify the carbon emission. 
In Table~\ref{tab_cv_results}, \tool consistently demonstrates remarkable efficiency across various model architectures. For the SimpleCNN model, while \tool incurs a 68.8\% rise in carbon emissions over the baseline, CAFL shows a substantial increase of 51.9\%. Similarly, SignSGD and JoPEQ trail behind with increments of 56.5\% and 26.0\%, respectively. However, as we scale to more intricate models, \tool's relative overhead diminishes, with only a 53.4\% increase for ResNet18. In contrast, methods like CAFL see their emissions skyrocket, with a staggering 5,789.6\% overhead for the ResNet18 model.
When juxtaposed with LDP, \tool's efficiency becomes more evident. While LDP boasts modest overheads for simpler models, its advantage diminishes with complex architectures. Additionally, CDPA provides a superior balance between environmental impact and model utility. Techniques such as GradDrop, Soteria, and SDQ, even though they bear lower privacy computational overheads, face challenges in mitigation of data reconstruction attacks.

\noindent \textbf{Model utility vs. computation \& communication overhead trade-off.~} illustrate the trade-off between model utility and computation \& communication overhead, we particularly report Figure~\ref{radars} based on Table~\ref{tab_cv_results} settings. We report top-1 model accuracy (Acc.), training efficiency (TE), carbon emission efficiency (CEE), and communication efficiency (CE) across various baseline methods, including gradient compression methods, compression \& privacy methods and privacy-preserving methods. Notably, \tool shows a superior trade-off. Among comparison with gradient compression methods (\ie signSGD, SDQ, and GradDrop), although signSGD demonstrates higher communication efficiency, its privacy protection is very limited. GradDrop has almost no privacy-preserving ability. Additionally, privacy-preserving methods (\ie LDP and Soteria) take up a substantial amount of communication bandwidth with suboptimal accuracy. Compression \& privacy methods (\ie JoPEQ and CAFL) encounter difficulties in Acc., TE and CEE. First, CAFL demands substantial carbon emissions to sustain its privacy-preserving capabilities, achieving only one-third the accuracy when compared to our \tool. Second, JoPEQ exhibits slightly lower TE and CEE relative to our \tool, while maintaining suboptimal accuracy. Note that other DP-based methods experience difficulty converging when $\epsilon < 2$, but $\epsilon = 1.38$ for \tool (\ie $p=0.8$) still maintain an acceptable and negligible accuracy loss(as demonstrated in Table~\ref{tab_various_epsilons} in Appendix). 

In addition, in Table \ref{tab_cv_results}, the accuracy column showcases the superiority of CDPA across these three models when benchmarked on the CIFAR-10 dataset. For the SimpleCNN and ConvNet models, the CDPA method either matches or surpasses the accuracy of other defensive techniques, with performances closely trailing the non-defensive approach. Remarkably, even in the more complex ConvNet and ResNet18 model, \tool stands out, registering an accuracy of 0.4\% increasing compared to the vanilla model, showcasing resilience against the significant drops observed in other techniques like CAFL and signSGD. Collectively, these data solidify the \tool method as a consistently high-performance strategy across different architecture sizes.
In summary, the significance of the \tool strategy in terms of utility is strongly emphasized, showcasing its outstanding performance under various conditions.

\subsection{Experimental Results in NLP Tasks}\label{sec_results_nlp}
In this section, we evaluate the performance of \tool in NLP tasks, with respect to model utility and defense performance. We did not estimate the communication cost and carbon emissions in NLP tasks, since the models are pre-trained, and the overhead may vary in practice.

\begin{figure}[t]
\centering
\includegraphics[width=\linewidth]{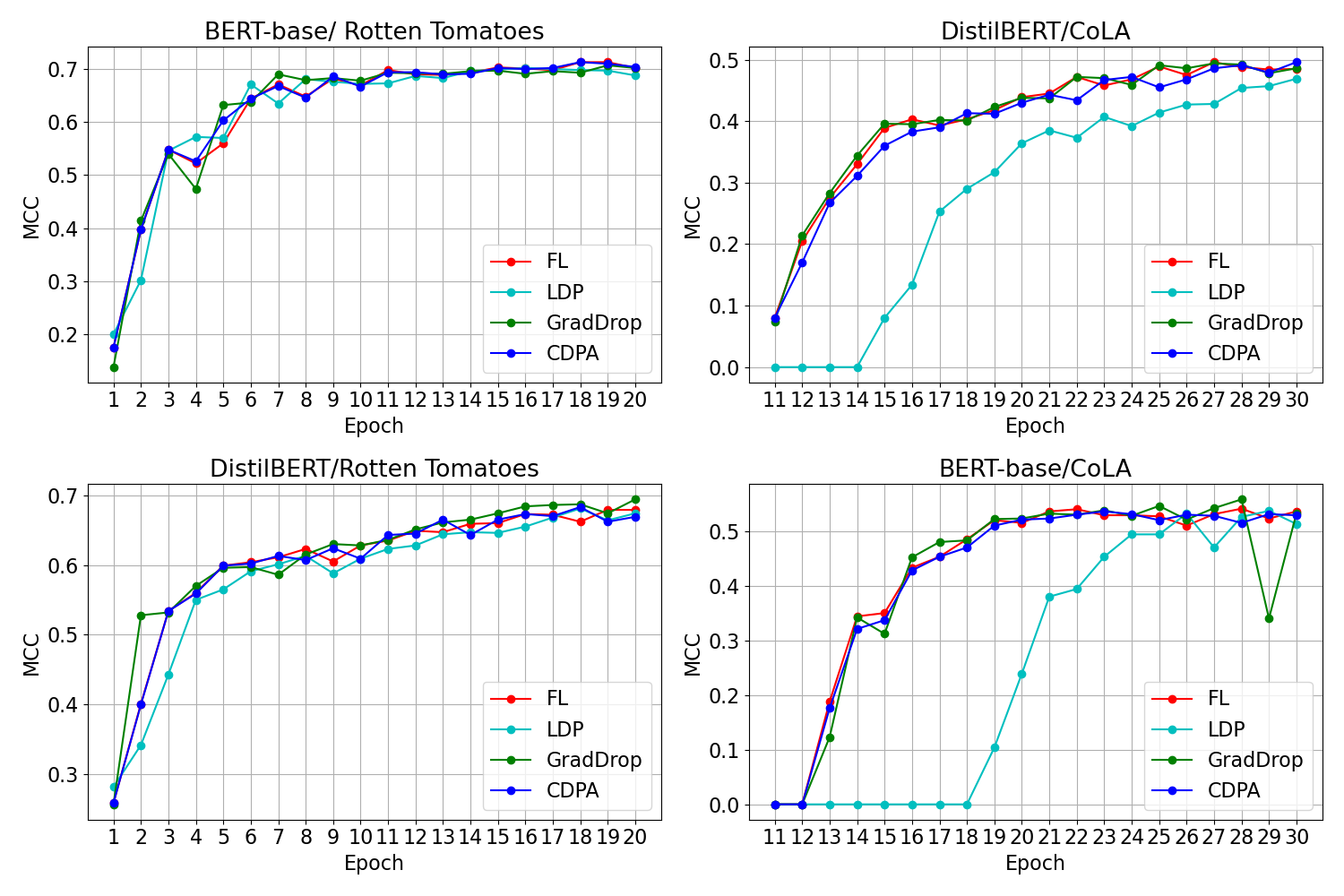}
\caption{Model utility performance in NLP tasks.}
\label{NLPutility}
\end{figure}

\noindent \textbf{Model utility in NLP tasks.~} 
Figure~\ref{NLPutility} delineates model utility for BERT\textsubscript{base} and BERT\textsubscript{distill} trained on Rotten Tomatoes and CoLA datasets. The performance curves indicate that \tool closely aligns with the vanilla model in utility, particularly evident by the BERT\textsubscript{base} and BERT\textsubscript{distill} models on Rotten Tomatoes dataset. Meanwhile, in the results on CoLA dataset, GradDrop demonstrates consistent utility performance, yet it does not consistently surpass \tool. Furthermore, LDP tends to underperform across all experiments. In summary, \tool emerges as a method that proficiently balances enhanced privacy with sustained model utility, rivaling or exceeding other contemporary techniques, underscoring its paramount significance in privacy-centric machine learning tasks.

\input{Tables/tab_nlpbase}

\input{Tables/tab_defense}
\noindent \textbf{Defense performance in NLP.~} 
We rigorously evaluate our defense method, \tool, against DLG, TAG, and LAMP attacks, focusing on models from the transformer architecture in NLP tasks. Transformer-based models are known for their vulnerabilities, particularly in the word embedding layer, where gradient analysis can reveal word usage in iterations. Additionally, the attention layer can leak information, making data reconstruction possible. To counter these challenges, \tool is applied on all layers within the transformer. 

As demonstrated in Table~\ref{nlpbase}, we compare defense performance across different benchmarks (\ie non-defense, GradDrop, and LDP) against DLG, TAG, and LAMP attacks in NLP tasks with CoLA dataset and Base BERT model. 
Notably, under the LAMP attack, \tool yielded an R-1 score of 19.78 and an R-2 score of 0.00, indicating a remarkable reduction in reconstructability compared to non-defense and other defense methods. Tables~\ref{defense2} offers a visual insight into the reconstructed examples resulting from the mentioned attacks. The reconstructed terms aligned with the ground truth are highlighted, showcasing the defense's strength. In the DLG, TAG, and LAMP attack scenarios, the reconstructions from our \tool method exhibit minimal to no alignment with the actual phrases, emphasizing its leakage-resilience. 
In summary, while the DP method exhibits some defensive potential by moderately reducing the ROUGH values, it still leaves room for key term reconstruction. On the other hand, our \tool method stands out by dramatically minimizing the chances of critical term reconstruction, solidifying its superior defensive prowess.

\section{Discussion \& Limitations}
In this section, we provide and summarize a more in-depth discussion based on experimental results. 

\noindent \textbf{Impact of model size.~} 
In our study, \tool demonstrates a remarkable advantage concerning model size. Unlike LDP, which tends to accumulate noise as the model size increases, resulting in a significant degradation of model utility or convergence rate, our \tool method exhibits impressive performance. In some instances, it even surpasses that of non-defense models, thereby breaking the conventional trade-off between privacy protection and utility. This exceptional performance can be attributed, in part, to the \tool's ability to maintain minimal reconstruction errors, effectively improving model generality to the model. This regularization mechanism contributes to improved model generalization, enabling it to perform exceptionally well on large-scale models without sacrificing its performance. As a result, \tool introduces a new paradigm in the relationship between privacy protection and model size, allowing researchers to flexibly control model scale without compromising crucial task utility. This groundbreaking advantage positions the \tool method as a promising solution for a wide range of tasks involving large models in the current landscape.

\noindent \textbf{Impact of gradient compression and DP on data reconstruction attacks.~} Many gradient compression techniques have inadvertently misled researchers into believing they effectively counter data reconstruction attacks. In our experiments, methods such as GradDrop and quantization (represented by SDQ and signSGD) struggled to prevent input reconstructions. This shortfall might be attributed to the inherent nature of these techniques, which primarily focus on efficient gradient representation rather than obfuscating gradient information sufficiently to thwart inversion attempts. In the previous researches, the most effective strategy against data reconstruction attacks, as observed, is the introduction of greater perturbations which introduces noise directly tailored to mask individual data points, making it inherently more resistant to data reconstruction attacks. However, Local Differential Privacy (LDP) may have limitations, especially when facing generative models with image priors. To address this concern, our bit-flipping method introduces a larger variance, mitigating these limitations and achieving a better trade-off between model utility and privacy preservation.


\noindent \textbf{Limitations.~}
We see two potential limitations in our research. First, critical infrastructure components may be located in remote or challenging environments with limited network connectivity, maintaining ACIs in such situations can be challenging. Without stable network connectivity, the slowest clients are at risk of being dropped from rounds of learning, which may lead to fairness issues. However, this issue has been widely discussed and addressed in recent studies~\cite{jiang2022pisces,nguyen2022federated,yu2023async,chai2020fedat}. The asynchronous federated learning solutions~\cite{jiang2022pisces,nguyen2022federated} can be adopted in ACI networks. Second, critical infrastructure networks may consist of a variety of devices and systems from different manufacturers and with varying data formats. Integrating and harmonizing data from these heterogeneous sources for federated learning can be complex. Several studies~\cite{huang2022learn,li2019fedmd,zhu2021data,fang2022robust} addressed the  heterogeneous client issue in federated learning with solutions such as transfer learning and knowledge distillation, which could be adoptive to ACIs. 

\section{Conclusion}
\label{conclusion}
We proposed \tool, a 
leakage-resilient, communication-efficient, and carbon-neutral aggregation approach towards featuring the federated AI-enabled Critical Infrastructure. Specifically, we designed a random bit-flipping strategy to safeguard the ACI network against potential data leakage. This strategy has been proven to be differentially private, while also making commendable efforts in energy saving and enabling the introduction of controllable noise with a larger variance through the design of the bit-flipping mask and the flipping probability.
Furthermore, the quantization mechanism developed in \tool halves the communication cost with negligible quantization error. We comprehensively evaluated the performance of the proposed \tool in both computer vision and natural language processing tasks, with respect to model utility, defense performance, communication cost, and carbon emissions. The experimental results indicate that \tool can effectively prevent data reconstruction against several state-of-the-art data reconstruction attacks, while preserving the model utility. 
We also demonstrated its superior adaptability over existing benchmarks. 
Notably, the minimized carbon emission positions \tool as a practical choice for real-world applications. We envision a roadmap where the inherent mechanisms of \tool can be fine-tuned and customized, catering to the unique demands of ACIs. This adaptability promises a future where privacy protection, resilience, and efficiency coexist harmoniously.

\bibliographystyle{IEEEtranS}
\bibliography{Reference_s}

\section*{Appendix}
\setcounter{section}{0}
\renewcommand{\thesection}{\Alph{section}}
\section{Federated Learning}\label{app_fl}
In federated learning, the machine learning task is distributed from a centralized server to a set of clients. Federated learning typically consists of two main components: local training on the client side and aggregation on the server side. 

\noindent \textbf{Local training at the client side.~} The client retrieves global updates, represented by $h_t$,  from the server and conducts a local training step on its private dataset by applying $h_{t+1} = h_{t} - \eta \Delta h_{t}$, where $\eta$ is learning rate. 

\noindent \textbf{Aggregation at the server side.~}
Based on receiving local updates from $R$ clients, the server incorporates these updates and update its global state, and initiates the next round of federated learning. The common aggregation frameworks are FedAvg~\cite{mcmahan2017communication} and FedSGD~\cite{mcmahan2017communication}. The difference of FedSGD is that the server aggregates the averaged gradients from the selected clients, while FedAvg aggregates the averaged weights from the selected clients. Note that when a client runs a local epoch, the aggregation results of FedAvg and FedSGD are the same. 

In our study, we adopt FedAvg as an alternative aggregation framework, especially for layers that are unselected and do not undergo bit-flipping operations.

\section{Subtractive Dithered Lattice Quantization(SDQ)}
\label{sdq_background}
Recent research~\cite{shlezinger2020uveqfed,shlezinger20221} indicates that SDQ outperforms QSGD~\cite{alistarh2017qsgd} in federated learning scenarios. 
This technique is designed to efficiently represent and transmit data while minimizing information loss. Unlike other existing quantization methods (\eg QSGD) that directly map data points to discrete levels, SDQ introduces a layer of randomness, known as dithering, into the quantization process.
In SDQ, dithering involves adding controlled noise or random variations to the input data prior to quantization. This noise helps break the symmetry of quantization errors, allowing for improved accuracy in representing data points, particularly in scenarios where fine-grained distinctions are essential.
One of the key advantages of SDQ is its ability to strike a balance between data compression and fidelity. By intelligently applying dithering, SDQ achieves compression without sacrificing the crucial details of the original data, making it particularly useful in applications where preserving data quality is paramount. Below, we outline the fundamental principles of SDQ.

In Lattice quantization~\cite{zamir1992universal}, continuous or high-dimensional data represented with $n$ bits is mapped to a lower-dimensional discrete set of points (\ie lattice), with $m$ bits.
Given the number of dimensions $L \in \mathcal{Z^+}$ and a non-singular $L \times L$ matrix $\mathbf{G}$, the discrete quantized function $Q_{\mathcal{L}}(\mathbf{x})$ will discover the nearest vector in the lattice $\mathcal{L}$: 
\begin{equation}
Q_{\mathcal{L}}(\mathbf{x}) = \arg \min_{\mathbf{p \in \mathcal{L}}}  \vert \vert \mathbf x - \mathbf p  \vert \vert_2,
\end{equation}
\begin{equation}
\text{where~}
\mathcal{L} \triangleq\left\{\boldsymbol{x}=\boldsymbol{G l}: \boldsymbol{l} \in \mathcal{Z}^L\right\}.
\end{equation}

In order to restrict the number of lattices within a particular bit range, a common practice is to confine the lattices within a sphere of radius $\gamma$. By partitioning the vector $\mathbf{x}$ to be quantized into multiple sub-vectors based on a given dimension $L$, such as $[\mathbf{x_1},...,\mathbf{x_n}]^T$, each sub-vector can be quantized independently to minimize the quantized error. 

Dithered lattice quantization (DQ)~\cite{lipshitz1992quantization} is a variant of lattice quantization. It involves adding a dither signal $\mathbf d$ to the sub-vectors before they are quantized. This dither signal is generated from a uniform distribution, which helps to improve the quantization error, which is because the random process of quantization distributes the quantization error more uniformly across the signal spectrum. The expression for dithered lattice quantization is:

\begin{equation}
Q_{\mathcal{L}}^{DQ}(\mathbf{x}) = Q_{\mathcal{L}}(\mathbf x + \mathbf d).
\end{equation}
Due to the distortion introduced by adding dither, subtractive dithered lattice quantization (SDQ)~\cite{gray1993dithered} improves the quantization performance by subtracting the dither signal from the DQ approach. The dither is subtracted out after the quantization process, which effectively reduces the overall quantization error. The key formula as follows:

\begin{equation}
Q_{\mathcal{L}}^{SDQ}(\mathbf{x}) =  Q_{\mathcal{L}}^{DQ}(\mathbf{x}) - \mathbf d.
\end{equation}

\section{Proof of Theorem 1} \label{apx:thm1}
To prove Theorem 1, we firstly prove Theorem 2, then Lemma 13 in~\cite{andriushchenko2022towards} should be satisfied under Assumption \textbf{(A4)}. We have the following
\begin{equation}
\mathbb{E} \langle \nabla L(\bm\theta, \bm\theta + \gamma \nabla L(\bm\theta)), \nabla L(\bm\theta) \rangle \geq (\frac{1}{2} - \hat \beta \gamma) \| \nabla L(\bm\theta) \|_2^2 - \frac{\hat \beta^2 \gamma {\sigma}^2}{2}.
\end{equation}
Then we can apply Theorem 15 in~\cite{andriushchenko2022towards}, then for all $\eta \leq \frac{1}{2\hat \beta}$ and $\gamma \leq \frac{1}{2\hat \beta}$, we have 
\begin{equation}
    \frac{1}{T} \sum_{t=0}^{T-1} \mathbb{E} \left\| \nabla L(\bm{\theta_t}) \right\|^2 \leq \frac{4\left( L(\bm{\theta_0}) - \mathbb{E} L(\bm{\theta_T}) \right)}{T \eta} + {4T  \sigma^2 \hat \beta^2(\eta + \gamma^2)},
    \label{111}
\end{equation}
where we substitutes Equation \ref{111} with PL-condition using Assumption \textbf{(A5)}, we have:
\begin{equation}
\begin{aligned}
\frac{1}{T} \sum_{t=1}^{T} \mathbb{E}[L_D(\bm{\theta_t})] - \inf_{\bm{\hat{\theta}}} L_D(\bm{\hat{\theta}}) 
&\leq \mu \cdot \left( \frac{4\left( L(\bm{\theta_0}) - \mathbb{E} L(\bm{\theta_T}) \right)}{T \eta} \right. \\
&\quad \left. + 4T {\sigma}^2 \hat{\beta}^2(\eta + \gamma^2) \right),
\end{aligned}
\label{eq:00}
\end{equation}
The proof concludes.
\section{Proof of Theorem 2} \label{apx:thm2}
To prove Theorem 2, we need to be under assumption \textbf{(A1)}, \textbf{(A2)} and \textbf{(A3)}. The full proof is adapted from \cite{wang2024dpadapter}
\begin{equation}
\begin{split}
\left| L_i(\bm{\theta}) - L_i(\bm{\theta}') \right| &= \left| L_i(f(x_i| \bm \theta), y_i) - L_i(f(x_i|{\bm{\theta}'}), y_i) \right| \\
&\leq \beta_1 \left| f(x_i| \bm \theta) - f(x_i|{\bm{\theta}'}) \right| \\
&\leq \beta_1 \| \bm{\theta} - \bm{\theta}' \|_2
\\
&= \xi \beta_1 \rho |1- p_{\gamma_2} | ,
\end{split}
\end{equation}
and another half result can be proved as follows:

\begin{equation}
\begin{split}
&\hspace{-0.5cm}\|\nabla L(\theta) - \nabla L(\theta')\|_2\notag \\
        &\hspace{-0.5cm}\ = \|\nabla_x L_i(f(x_i| \theta), y_i) \nabla f(x_i| \theta) - \nabla_x L_i(f(x_i| \theta'), y_i) \nabla f(x_i|\theta')\|_2\notag \\
        &\hspace{-0.5cm} \leq \|\nabla_x L_i(f(x_i| \theta), y_i) \nabla f(x_i| \theta) - \nabla_x L_i(f(x_i| \theta'), y_i) \nabla f(x_i|\theta)\|_2\notag + \\
        &\hspace{-0.5cm} \|\nabla_x L_i(f(x_i| \theta'), y_i) \nabla f(x_i| \theta) - \nabla_x L_i(f(x_i| \theta'), y_i) \nabla f(x_i|\theta')\|_2\notag \\
        & \hspace{-0.5cm}\leq \|\nabla f(x_i|\theta)\|_2|\nabla_x L_i(f(x_i|{\theta}), y_i) - \nabla_x L_i(f(x_i|{\theta'}), y_i)| + \\
        & |\nabla_x L_i(f(x_i|{\theta'}), y_i)|\|L_i f(x_i|{\theta}) - \nabla f(x_i|{\theta'})\|_2\notag \\
        & \hspace{-0.5cm}\leq \rho^2\beta_2\|\theta - \theta'\|_2 + \beta_1\beta\|\theta - \theta'\|_2 \\
        & \hspace{-0.5cm}= (\rho  \beta_2 + \beta_1) \| \xi \left(1- p_{\gamma_2} \right)\|_2 
\end{split}
\end{equation}

Finally, the proof concludes.

\section{Survey Protocol}
\label{app_survey_protocol}

To ensure the fairness of the survey results, we randomly selected 10 images representing the reconstruction results of different methods to be included in the survey. Consequently, our survey comprised 60 questions in total, corresponding to the six defense schemes. Participants were tasked with assessing the similarity of two presented images based on a Likert scale ranging from -2 to 2, representing ``Very unlike'', ``Unlike'', ``Unsure'', ``Like'', and ``Very like'' respectively. The entire set of 60 questions was designed to be completed in 15 minutes.

We engaged 100 native English-speaking participants through the Amazon Mechanical Turk platform, where approximately 57\% identified as female. All participants were above the age of 18, predominantly hailing from the United States, the United Kingdom, and Australia. We maintained an inclusive approach, without specific restrictions on educational background or professional experience, resulting in a participant distribution aligned with the platform's average profile. We paid each participant at a rate of \$12.00 USD per hour, which is above the average payment rate in the United States. We followed best practices for ethical human subjects survey research, e.g., all questions were optional, and we did not collect unnecessary personal information. All participants are over 18 years old, and they consented for their answers to be used for academic research. The data are stored on a secure server approved by the Ethics Committee and will be removed after the study is completed.

\begin{figure}[t]
\centering
\includegraphics[width=\linewidth]{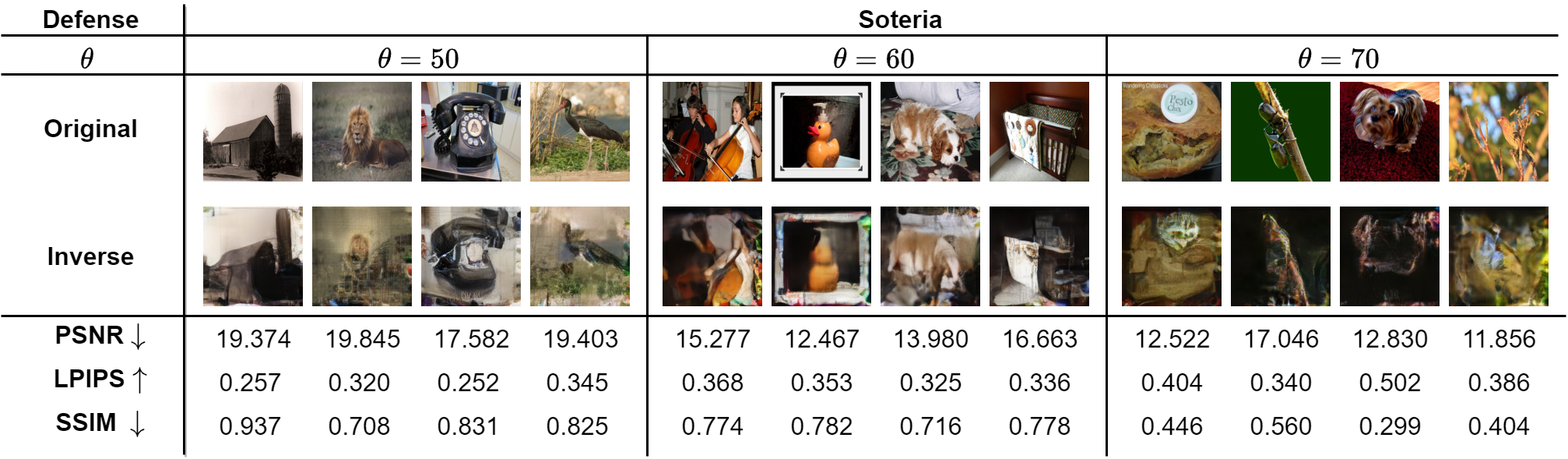}
\caption{Soteria against ROG attacks.}
\label{fig_soteria}
\end{figure}

\begin{figure}[t]
\centering
\includegraphics[width=\linewidth]{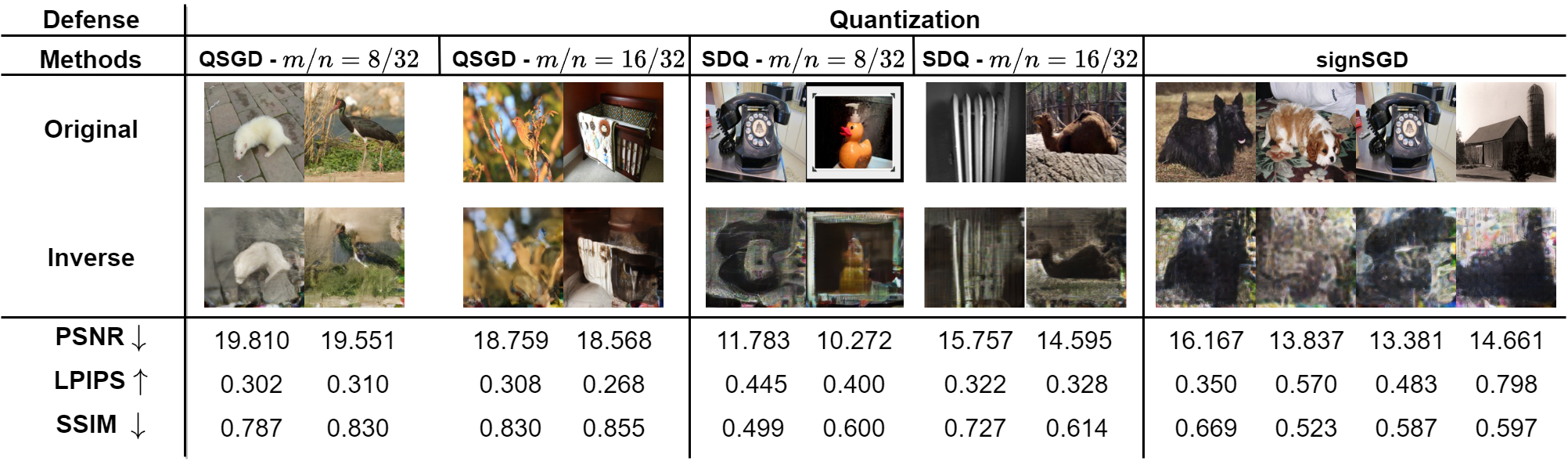}
\caption{Quantization against ROG attacks.}
\label{fig_quantization}
\end{figure}

\begin{figure}[t]
\centering
\includegraphics[width=\linewidth]{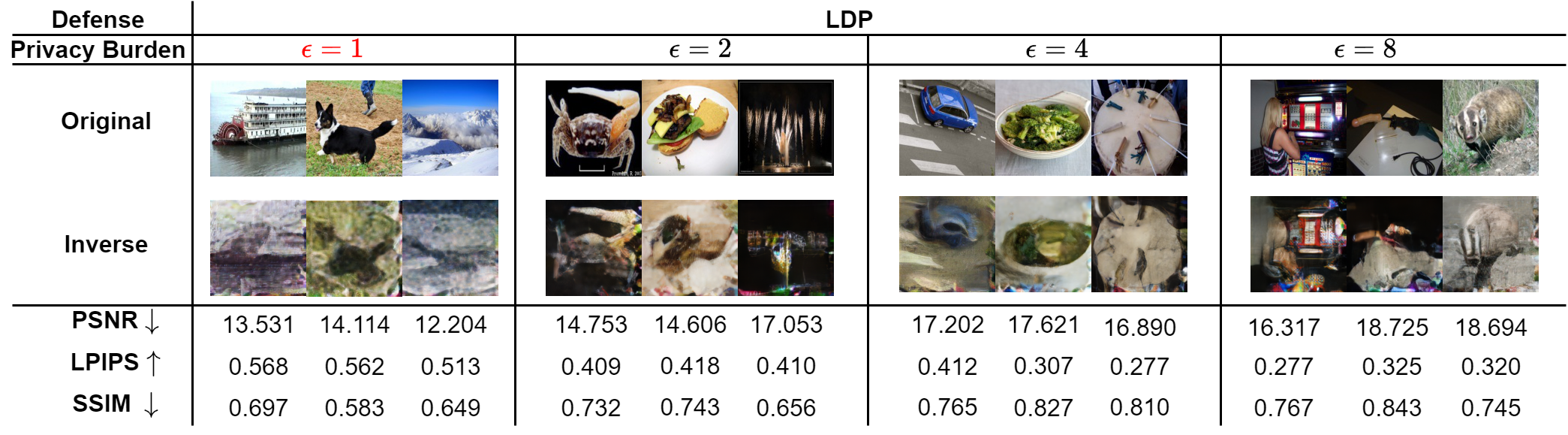}
\caption{LDP against ROG attacks. The red color signifies that under this setting, the model can not achieve convergence. }
\label{fig_LDP}
\end{figure}

\begin{figure}[t]
\centering
\includegraphics[width=\linewidth]{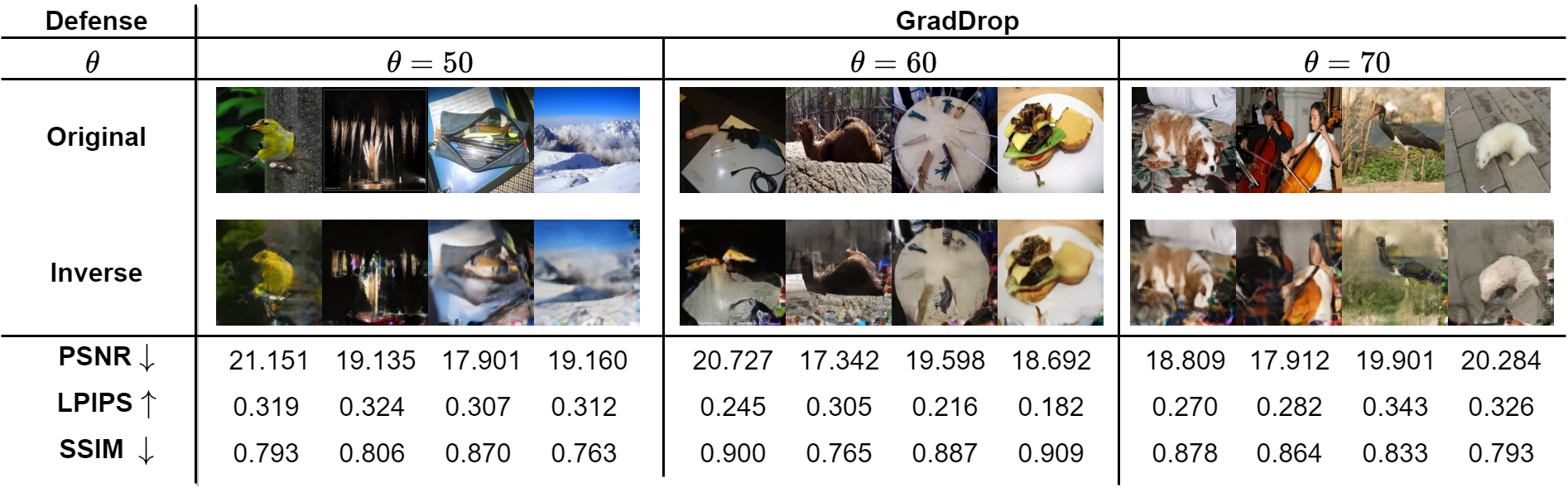}
\caption{GradDrop against ROG attacks.}
\label{fig_GradDrop}
\end{figure}
\begin{figure}[t]
\centering
\includegraphics[width=\linewidth]{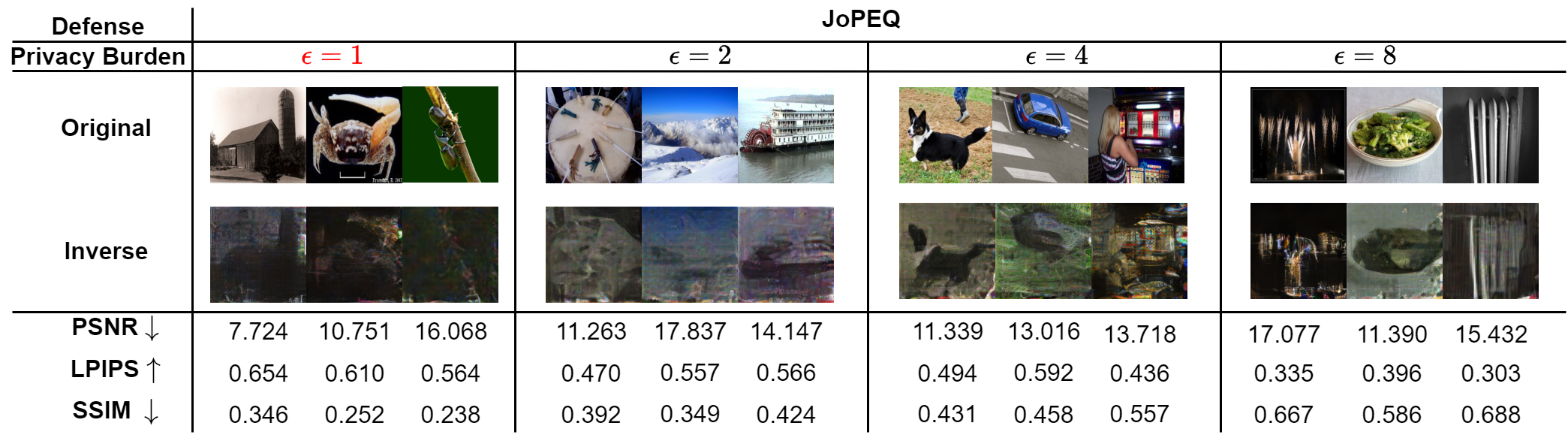}
\caption{JoPEQ against ROG attacks. 
The red color signifies that under this setting, the model can not achieve convergence.  }
\label{fig_jopeq}
\end{figure}
\begin{figure}[t]
\centering
\includegraphics[width=\linewidth]{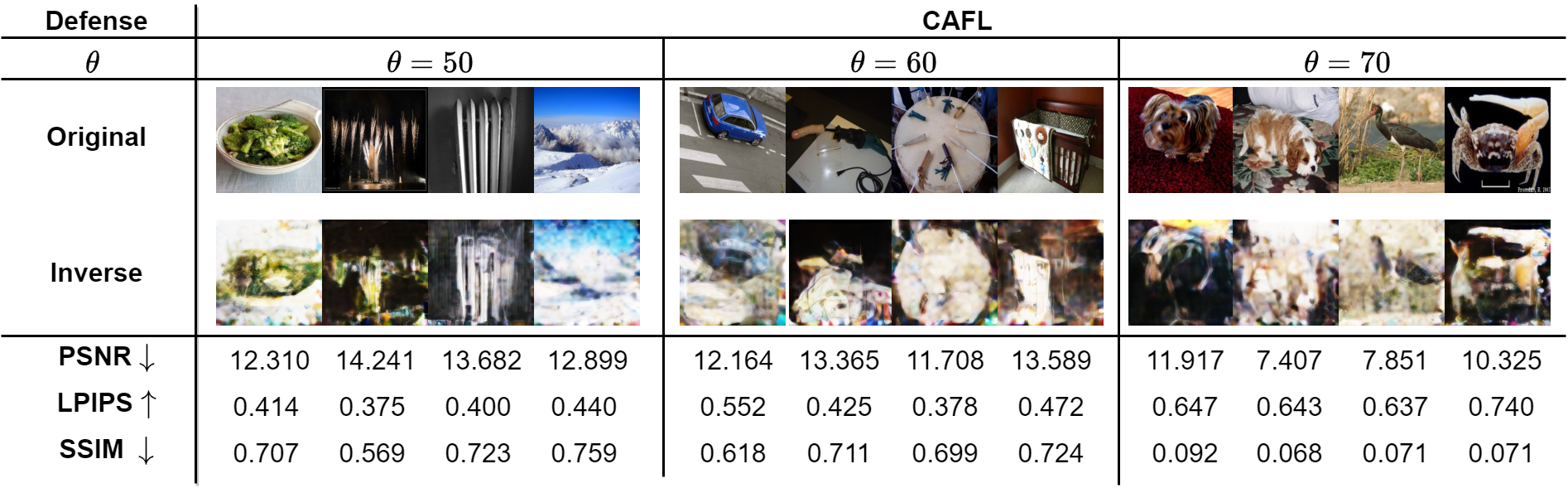}
\caption{CAFL against ROG attacks.}
\label{fig_cafl}
\end{figure}
\section{Additional Experiments against ROG} \label{app_extra}
In this section, we carry out additional supplementary experiments with extensive hyperparameter configurations for ROG attacks, presenting further results that complement those provided in Section 6.1 
and Figure 6, 
Not limited to defense methods, our expanded reconstructed results also encompass communication-efficient methods, including QSGD, SDQ, and signSGD. The results are shown in Figures~\ref{fig_soteria}, ~\ref{fig_quantization},~\ref{fig_LDP},~\ref{fig_GradDrop},~\ref{fig_jopeq} and ~\ref{fig_cafl}.


\section{Data Reconstruction from Bit-flipping}
We further considered a new attack scenario in which the attacker is strong enough to identify flipped gradients through outlier detection and further impute the outliers (\ie the flipped gradients) with the mean of gradients or simply set them to 0. Additional experiments are conducted on the Imagenet dataset with the ROG attack. As shown in Table~\ref{tab_reconstruction}, CDPA is robust against such attacks.

\input{Tables/tab_reconstruction_from_flipping}

\section{Differential Privacy}\label{app_dp}
Differential privacy~\cite{dwork2008differential} is a mathematical framework that allows for the analysis of sensitive data while ensuring the privacy of individuals in the dataset.

\noindent\textbf{Definition 1} (($\epsilon,\delta$) - Differntial Privacy(DP)). $\mathcal{A}$ represents the randomized algorithm used to analyze the data, given $D$ and $D'$ that represent two datasets that differ by only one individual's data, if they have the guarantee~\cite{dwork2008differential}:  

\begin{equation}
\Pr[\mathcal{A}(D) \in S] \leq e^{\epsilon} \cdot \Pr[\mathcal{A}(D') \in S] + \delta \label{DP}
\end{equation}
where $\epsilon$ represents the level of privacy protection provided by the algorithm (privacy budget), $\delta$ is the failure probability, and $S$ is the set of possible outputs of the algorithm. 

The privacy budget $\epsilon$ refers to the level of encryption applied to data, where a smaller privacy budget signifies stronger privacy protection. The $\delta$ represents the probability that algorithm $\mathcal{A}$ fails to satisfy $\epsilon-DP$, providing a certain level of fault tolerance to the entire formula (\ref{DP}). 

In this work, we specifically examine the use of Local Differential Privacy (LDP) with the Laplace mechanism~\cite{dwork2014algorithmic} to upper bound privacy leakage, particularly when transmitting gradient information from the client to the server.
\section{Experiment for p-selection}\label{p_selection}
Additional experimental results with varying $p$ values and corresponding $\epsilon$ values are provided in Table~\ref{tab_various_epsilons}. A higher $p$ may lead to higher accuracy but lower defense performance against reconstruction attacks. However, when $p$ is selected within a ideal range that will not induce high recovery error (e.g., $0.8 < p < 1$ shown in Figure 4 
), its influence on model accuracy and defense performance is not significant.

\input{Tables/tab_various_epsilons}
\section{Attaks in Experiments}
\label{attaks_exp}
\begin{itemize}[leftmargin=*]
\item \textbf{DLG~\cite{zhu2019deep}.} The malicious server refines reconstructed data to minimize the Euclidean distance between original and reconstructed gradients in backpropagation. We use this attack in both CV and NLP experiments. 

\item \textbf{IG~\cite{geiping2020inverting}.} Unlike DLG, this method reconstructs private images by optimizing cosine similarity instead of Euclidean distance. We apply this attack in CV experiments. 
\item \textbf{ROG~\cite{yue2023gradient}. }This method leverages a trained GAN and post-processing technologies like denoising and super-resolution to enhance the reconstruction of private data from the latent space. We apply this attack in CV experiments.
\item \textbf{TAG~\cite{deng2021tag}.} For the first time, the data reconstruction attack is being applied to transformer-based language models in the field of NLP. This approach not only employs Euclidean distance but also adopts the L1 norm in reconstruction loss to tackle adaptability issues within these models.
\item \textbf{LAMP~\cite{balunovic2022lamp}.} LAMP adopt an auxiliary language model with prior knowledge to guide the reconstructed embeddings towards more natural text. Additionally, LAMP also switches between continuous and discrete optimization to lessen the loss when reconstructing embeddings. To avoid getting stuck at local minima, it performs discrete text transformations. We apply this attack in NLP experiments. 
\end{itemize}
\section{Defense Benchmarks}\label{defense_benchmarks}
\begin{itemize}[leftmargin=*]
\item \textbf{LDP~\cite{wei2020federated}} adds a specific amount of noise to the uploaded gradients to prevent malicious users from adopting data reconstruction attacks using the gradients.
\item \textbf{Soteria~\cite{sun2021soteria}} selects a fully connected layer within the convolutional neural network model to defend and introduces perturbations to the data representation.
\item \textbf{SignSGD~\cite{bernstein2018signsgd,bernstein2018signsgd2}} discretizes the transmitted signal using only one bit, which is determined by the sign of the signal. The method assumes that the sign of the signal is the dominant factor in determining the gradient descent during data transmission.
\item \textbf{GradDrop~\cite{aji2017sparse}} prunes a part of gradients based on their absolute values, leading to fewer parameter updates.
\item \textbf{SDQ~\cite{shlezinger2020uveqfed}} is a vector quantization method that reduces quantization error by employing random dithering and subtraction.
\item \textbf{CAFL~\cite{miao2022compressed}} combines Compressive Sensing and local differential privacy. In other words, it employs Compressive Sensing to reduce the number of parameters, while using LDP to enhance privacy preservation.
\item \textbf{JoPEQ~\cite{lang2023joint}} merges SDQ and local differential privacy (LDP). By applying SDQ with a quantization rate of $a$, it lowers communication costs, while leveraging LDP to bolster privacy protection.
\end{itemize}
\section{Datasets in Experiments}\label{datasets}
\begin{itemize}[leftmargin=*]
\item \textbf{CIFAR10~\cite{krizhevsky2009learning}} is an image classification dataset that consists of 10 different categories, each containing 6,000 color images of size 32x32 pixels. 
We use this dataset in all utility and defense performance evaluations.

\item \textbf{ImageNet~\cite{deng2009imagenet}} comprises over 14 million labeled images, distributed across more than 20,000 categories or ``synsets''. Each image in the dataset is annotated with a specific class label, making it a vital resource for the supervised learning domain. We use this dataset in ROG attacks, but not in DLG and IG attacks, because the complexity and diversity of ImageNet images make DLG and IG attacks less effective, diminishing their feasibility and impact in real-world scenarios.
\item \textbf{CoLA~\cite{warstadt2019neural}}. The CoLA dataset comprises sentences made up of 5-9 words, with the task being to label whether the grammar of the sentence is incorrect. 
We use this dataset in all utility and defense performance evaluations.
\item \textbf{RottenTomatoes~\cite{pang2005seeing}}. The Rotten Tomatoes dataset consists of 14-27 words per sentence, tasked with annotating movie reviews as either positive or negative in sentiment. BERT\textsubscript{base} and BERT\textsubscript{distill} are also trained on this dataset. 
\end{itemize}
\section{Metrics in CV Tasks}\label{app_metrics}
\noindent \textbf{Model utility metrics.~} In CV tasks, we measure the utility of the models by evaluating the Top-1 accuracy and loss of the global mode. In NLP tasks, we use Matthews Correlation Coefficient (MCC)~\cite{chicco2020advantages} as a model utility metric, which ensures a comprehensive and balanced assessment of model performance, accounting for both true and false positives and negatives, regardless of dataset imbalance. The higher MCC indicates a higher model performance.

\noindent \textbf{Communication cost metrics.~} Communication cost refers to the average number of model updates occupied by clients when transmitting to the server. The parameter quantity is measured in megabits (Mb), representing the size of data uploaded per transmission epoch.

\noindent \textbf{Defense performance metrics.~} 
In CV tasks, the reconstructed image quality metrics include PSNR~\cite{castleman1996digital}, SSIM~\cite{wang2004image}, and LPIPS~\cite{zhang2018unreasonable}. These metrics are widely used to statistically measure the similarity between the original and reconstructed images. More details are provided in Appendix~\ref{app_metrics}.
In NLP task, we apply ROUGE family~\cite{lin2004rouge}, ROUGE-1, ROUGE-2 and ROUGE-L to measure the number of identified unigrams, the count of recognized bigrams, and the proportion of the length of the longest matching subsequence relative to the entire sequence's length.

\noindent \textbf{Carbon emission estimation.~} We estimate the carbon emission of each model based on its training time. Particularly, the carbon emission is estimated as CO2-equivalents (CO2eq) per 100 training epoch, using 
{Machine Learning Impact Calculator}~\cite{lacoste2019quantifying}, which has been used in several recent research~\cite{killamsetty2021retrieve, kuenneth2023polybert,dumontmodular}. 

In CV tasks, the reconstructed image quality metrics include PSNR~\cite{castleman1996digital}, SSIM~\cite{wang2004image}, and LPIPS~\cite{zhang2018unreasonable}. These metrics are widely used to statistically measure the similarity between the original and reconstructed images, which are defined as follows.

\begin{itemize}[leftmargin=*]
\item \textbf{PSNR~\cite{castleman1996digital}.~} PSNR quantifies the ratio between the maximum potential power of a signal and the power of the distorting noise that impacts the quality of its representation, which is given by:
\begin{equation}
\text{PSNR} = 10 \cdot \log_{10} \left( \frac{\text{MAX}_I^2}{\text{MSE}} \right), 
 \label{psnr}
\end{equation}
where $MAX$ is the maximum pixel value of an image, and $MSE$ is the Mean Square Error of two images, which could be defined as following:
\begin{equation}
\text{MSE}(I_1, I_2) = \frac{1}{a \times b} \sum_{i=1}^{a} \sum_{j=1}^{b} (I_1(i,j) - I_2(i,j))^2, \label{mse}
\end{equation} 
where $i,j$ indicate the position of pixels.
A higher value of PSNR indicates a higher similarity. 
\item \textbf{SSIM~\cite{wang2004image}.} SSIM Index is an image quality metric, which is defined as:
\begin{equation}
\text{SSIM}(I_1, I_2) = \frac{(2\mu_{I_1}\mu_{I_2} + c_1)(2\sigma_{I_1,I_2} + c_2)}{({\mu_{I_1}}^2 + {\mu_{I_2}}^2 + c_1)({\sigma_{I_1}}^2 + {\sigma_{I_2}}^2 + c_2)}, \label{ssim}
\end{equation} 
where $c_1$ and $c_2$ is small constants, $\mu_{I_1}$ and $\mu_{I_2}$ are the means of images $I_1$ and $I_2$, separately, ${\sigma_{I_1}}^2$ and ${\sigma_{I_2}}^2$ are the variance of $I_1$ and $I_2$, $\sigma_{I_1,I_2}$ is the covariance. A higher value of SSIM means a higher perception similarity. 
\item \textbf{LPIPS~\cite{zhang2018unreasonable}.~}
LPIPS evaluate the distance between image patches, which is defined as:
\begin{equation}
\text{LPIPS}(I_1, I_2) = \sqrt{\sum_{i} w_i \left(f_i(I_1) - f_i(I_2)\right)^2},
\label{lpips}
\end{equation} where $f_i(I)$ is the feature representation of $i$-th layer on VGG network. A higher value of SSIM means a lower perception similarity. 
\end{itemize}

\section{Experiment Settings}\label{app_ex}
In this section, we summarize and outline all experiment settings. Unless otherwise specified, the experimental results for other schemes are derived from the same hyperparameter configurations.

\noindent \textbf{CV Experiments:~}
\begin{itemize}[leftmargin=*]
\item \textbf{Defense Experiments}. For DLG attacks, we utilize the L-BFGS optimizer, while for IG attacks, we use the Adam optimizer with a learning rate of 0.1. The reconstructed images for both attacks are obtained after 300 iterations. For the CDPA scheme, we set probability $p = 0.98$, $m/n = 16/32$, $z = 4$ and flip position $I \in \{2,3\}$. For ROG attack, we typically set batch size to 16, federated learning rate is 0.003 and ROG learning rate is 0.05. In Figure 6, 
we use $\epsilon = 1$ in LDP and CAFL, and $\epsilon = 2$ in JoPEQ, $\theta_{GradDrop} = 0.6$ in FL, $\theta_{soteria} = 0.6$ in Soteria and $\theta_{GradDrop} = 0.5$ in CAFL.
\item \textbf{Utility Experiments}. The client count is 20 and the global epoch is set to 95. We use SGD optimizer with 0.1 learning rate. Local iteration is set to 100. The training batch size is 16. In Table 4, 
we set $\theta_{CS}$ = 0.5 \& $\epsilon = 2$ in CAFL; $\theta_{GradDrop}$ = 0.6 in GradDrop; $\theta_{Soteria}$ = 0.6 in Soteria; $a$ = 16 in SDQ.
\end{itemize}
\noindent \textbf{NLP Experiments:~}
\begin{itemize}[leftmargin=*]
\item \textbf{Defense Experiments}. For CDPA, we set the probability $p = 0.999$, $z = 5$ and the flip position $I \in \{5,6\}$. In Table 6, 
we set $\theta_{GradDrop} = 0.2$ in GradDrop and $noise = 0.005$ in DP. 
\item \textbf{Utility Experiments}. Unlike CV tasks, the number of clients in NLP tasks is set to 4. We use a pre-trained model that has already trained on WikiPedia~\cite{wikimedia2023} and BookCorpus~\cite{zhu2015aligning} datasets. We set global epoch to 20 on Rotten Tomatoes datasets and 30 on CoLA datasets. For defense method. we use the same setting as defense experiments. 
\end{itemize}

\end{document}

%% file: _header.tex
\usepackage{hyperref}
\usepackage{tikz}
\usepackage{amsmath}
\usepackage{bm}
\usepackage{graphicx}
\usepackage{multirow}
\usepackage{subcaption} 
\usepackage[font=small,labelfont=bf]{caption} 
\usepackage{footnote}
\usepackage{listings}
\usepackage{xspace}
\usepackage{booktabs}
\usepackage[ruled,vlined,linesnumbered]{algorithm2e}
\SetKwComment{Comment}{$\triangleright$\ }{}
\SetKwInOut{Variables}{Variables}
\usepackage{algorithm2e}
\usepackage{soul}
\usepackage{algorithmicx}  

\usepackage{algpseudocode}  
\usepackage{amssymb}
\usepackage[flushleft]{threeparttable}
\usepackage{xcolor}
\usepackage{lipsum}
\usepackage{caption} 
\usepackage{enumitem}
\usepackage{graphicx}

\setlist{noitemsep}

\newtheorem{theorem}{Theorem}

\usepackage{url}
\def\UrlBreaks{\do\A\do\B\do\C\do\D\do\E\do\F\do\G\do\H\do\I\do\J
\do\K\do\L\do\M\do\N\do\O\do\P\do\Q\do\R\do\S\do\T\do\U\do\V
\do\W\do\X\do\Y\do\Z\do\[\do\\\do\]\do\^\do\_\do\`\do\a\do\b
\do\c\do\d\do\e\do\f\do\g\do\h\do\i\do\j\do\k\do\l\do\m\do\n
\do\o\do\p\do\q\do\r\do\s\do\t\do\u\do\v\do\w\do\x\do\y\do\z
\do\.\do\@\do\\\do\/\do\!\do\_\do\ \do\;\do\>\do\]\do\)\do\,
\do\?\do\'\do+\do\=\do\#}

\renewcommand{\tabcolsep}{5pt}

\definecolor{firebrick}{rgb}{0.7, 0.13, 0.13}
\definecolor{darkblue}{rgb}{0,0,0.55}
\definecolor{grey}{rgb}{0.8,0.8,0.8}

\newcommand{\one}{({\em i}\/)}
\newcommand{\two}{({\em ii}\/)}
\newcommand{\three}{({\em iii}\/)}



\usepackage{siunitx}
\newcommand{\pie}[1]{%
\begin{tikzpicture}
 \draw (0ex,0ex) circle (1ex);
 \fill (0ex,-1ex) arc (-90:(#1-90):1ex) -- (0ex,-1ex) -- cycle;
\end{tikzpicture}%
}

\newcommand{\rect}[1]{%
\begin{tikzpicture}
 \draw (-1ex,-1ex) rectangle (1ex,1ex);
 \fill (#1ex,-1ex) rectangle (1ex,1ex);
\end{tikzpicture}%
}

\def\eg{\emph{e.g.,}\xspace}

\def\ie{\emph{i.e.,}\xspace}
\def\etal{\emph{et al.}\xspace}
\def\vs{\emph{vs.}\xspace}

\newcommand{\zehang}[1]{{\textcolor{orange}{#1}}}

\definecolor{codegreen}{rgb}{0,0.6,0}
\definecolor{codegray}{rgb}{0.5,0.5,0.5}
\definecolor{codepurple}{rgb}{0.58,0,0.82}
\definecolor{backcolour}{rgb}{0.95,0.95,0.92}
\newcommand{\up}{\textcolor{codegreen}{$\uparrow$}}
\newcommand{\down}{\textcolor{red}{$\downarrow$}}

\newcommand{\tool}{\textsc{CDPA}\xspace}
\newcommand{\toollong}{\textbf{C}ompressed \textbf{D}ifferentially \textbf{P}rivate \textbf{A}ggregation\xspace}

\SetKwInOut{KwData}{Data}
\SetKwInOut{KwIni}{Initialization}
\SetKwInOut{KwEn}{Encoding}
\SetKwInOut{KwDe}{Decoding}

%% file: _data.tex
\DeclareMathOperator{\sdq}{\texttt{SDQ}}
\DeclareMathOperator{\secureAddition}{\texttt{secure\_bin\_addition}}
\DeclareMathOperator{\floatToBin}{\texttt{float\_to\_bin}}
\DeclareMathOperator{\isMasked}{\texttt{is\_masked}}
\DeclareMathOperator{\bitFlip}{\texttt{bit\_flip}}
\DeclareMathOperator{\sendToServer}{\texttt{send\_to\_server}}
\DeclareMathOperator{\recover}{\texttt{recover}}
\DeclareMathOperator{\binToFloat}{\texttt{bin\_to\_float}}

%% file: Tables/tab_related_work.tex
\begin{table*}[t]
\centering
\caption{Existing mitigation against data leakage risks and/or communication bottleneck that can be used in ACI networks.}\label{tab_related_work}
\resizebox{\linewidth}{!}{
\begin{threeparttable}
\begin{tabular}{lccccccccc}
\toprule
\multirow{3}{*}{\textbf{Methods}} & 
\multicolumn{5}{c}{\textbf{Technology}} &
\multicolumn{2}{c}{\textbf{Mitigation against}} &
\multicolumn{2}{c}{\textbf{Application}} \\
\cmidrule(l){2-6}\cmidrule(l){7-8}\cmidrule(l){9-10}
 &
\textbf{Pruning} &
\textbf{Compressive Sensing} &
\textbf{Quantization} &
\textbf{DP} &
\textbf{Bit-flipping} &
\textbf{\begin{tabular}[c]{@{}c@{}} Communication Cost\end{tabular}} & 
\textbf{\begin{tabular}[c]{@{}c@{}} Data Reconstruction\end{tabular}} & 
\textbf{\begin{tabular}[c]{@{}c@{}}CV  Task\end{tabular}} & 
\textbf{\begin{tabular}[c]{@{}c@{}}NLP  Task\end{tabular}} \\
\midrule
signSGD~\cite{bernstein2018signsgd,bernstein2018signsgd2} & \pie{0} & \pie{0} & \pie{360} & \pie{0} & \pie{0} & \pie{360} & \pie{0} & \rect{-1} & \rect{1} \\ 
GradDrop~\cite{aji2017sparse} & \pie{360} & \pie{0} & \pie{0} & \pie{0} & \pie{0} & \pie{360} & \pie{0} & \rect{-1} & \rect{-1}  \\ 
SDQ~\cite{shlezinger2020uveqfed} & \pie{0} & \pie{0} & \pie{360} & \pie{0} & \pie{0} & \pie{360} & \pie{0} & \rect{-1} & \rect{1}  \\
Soteria~\cite{sun2021soteria} & \pie{360} & \pie{0} & \pie{0} & \pie{0} & \pie{0} & \pie{0} & \pie{360} & \rect{-1} & \rect{1}  \\ 
LDP~\cite{wei2020federated} & \pie{0} & \pie{0} & \pie{0} & \pie{360} & \pie{0} & \pie{0} & \pie{360} & \rect{-1} & \rect{-1}  \\ 
CAFL~\cite{miao2022compressed} & \pie{0} & \pie{360} & \pie{0} & \pie{360} & \pie{0} & \pie{360} & \pie{360} & \rect{-1} & \rect{1}  \\ 
JoPEQ~\cite{lang2023joint} & \pie{0} & \pie{0} & \pie{360} & \pie{360} & \pie{0} & \pie{360} & \pie{360} & \rect{-1} & \rect{1} \\ 
\midrule
\textbf{\tool (ours)} & \pie{0} & \pie{0} & \pie{360} & \pie{360} & \pie{360} & \pie{360} & \pie{360} & \rect{-1} & \rect{-1} \\
\bottomrule
\end{tabular}
\begin{tablenotes}
\item[]\pie{360} (\pie{0}): the method is (not) designed to mitigate the specific challenge, or the specific technology is (not) used in the method; \rect{-1} (\rect{1}): The method is (not) evidenced to be applicable in the specific machine learning task.
\end{tablenotes}
\end{threeparttable}
}
\end{table*}

%% file: Tables/tab_generalization_error.tex
\begin{table}[t]
\centering
\caption{Regularization effect introduced by CDPA.}
\label{tab_generalization_error}
\resizebox{\linewidth}{!}{
\begin{tabular}{@{}ccccc@{}}
\toprule
 & Vanilla model & $p=0.90$ & $p=0.95$ & $p = 0.98$ \\ 
\midrule
Accuracy & 79.4 & 79.2 & 79.4 & 79.8 \\
Generalization Error & 0.78 & 0.56 & 0.66 & 0.67 \\ \bottomrule
\end{tabular}
}

\end{table}

%% file: Tables/tab_defense_ROG.tex
\begin{table}[t]
\centering
\caption{Defensive performance against ROG attack, using Resnet18 model on ImageNet dataset. }
\label{tab_defense_ROG}
\resizebox{0.7\linewidth}{!}{
\begin{tabular}{lrrrr}
\toprule
Defenses & PSNR $\downarrow$ & SSIM $\downarrow$ & LPIPS $\uparrow$ \\
\midrule
Non-defense& 21.701 & 0.923 & 0.186\\
GradDrop& 18.656 & 0.829 & 0.346 \\
CAFL & 10.715 & 0.256 & 0.649 \\
JoPEQ & 12.203 & 0.416 & 0.561 \\
LDP & 14.382 & 0.564 & 0.570 \\
Soteria & 15.408 & 0.679 & 0.440 \\
Ours (CDPA) & \textbf{5.114} & \textbf{0.190} & \textbf{0.732} \\
\bottomrule
\end{tabular}
}
\end{table}

%% file: Tables/tab_privacy_overhead.tex
\begin{table}[t]
\centering
\caption{A comparison of performance with respect to accuracy, carbon emission, and communication overhead, across different benchmark methods on CIFAR-10 dataset. Note that we set $\epsilon = 4 $ on both LDP and JoPEQ as $\epsilon$ of CDPA is approximately 3.89.}\label{tab_cv_results}
\resizebox{\linewidth}{!}{
\begin{threeparttable}
\begin{tabular}{llrrrr}
\toprule
Models & Methods & Accuracy (\%)
& \begin{tabular}[c]{@{}c@{}}Training\\Time (min)\end{tabular}
& \begin{tabular}[c]{@{}c@{}}Carbon Emission\\(kg CO2)\end{tabular}
& \begin{tabular}[c]{@{}c@{}}Average \\ Communication Cost\\(Mb)\end{tabular}\\
\midrule
\multicolumn{1}{c}{\multirow{10}{*}{SimpleCNN}} 
 & Non-defense & 54.9 (-) & 23 & 7.7 (-) & 0.0087 (-) \\ 
 & CAFL & 45.2 (8.8\down) & 35 & 11.7 (51.9\%\up) & 0.0067 (23.0\%\down) \\ 
 & SignSGD & 51.9 (3.0\down) & 36 & 12.1 (56.5\%\up) & 0.0004 (95.4\%\down) \\ 
 & JoPEQ & 52.6 (2.3\down) & 29 & 17.4 (26.0\%\up) & 0.0041 (52.8\%\down)  \\ 
 & GradDrop & 53.3 (1.5\down) & 26 & 8.7 (13.0\%\up) & 0.0036 (58.6\%\down) \\ 
 & Soteria & 54.8 (0.1\down) & 43 & 14.4 (86.9\%\up) & 0.0044 (49.4\%\down) \\ 
 & SDQ & 54.8 (0.1\down) & 26 & 8.7 (13.0\%\up) & 0.0041 (52.8\%\down) \\ 
 & LDP & 52.8 (2.1\down) & 26 & 8.7 (13.0\%\up) & 0.0087 (0.0\%) \\ 
 & Ours & \textbf{54.8} (\textbf{0.1\down}) & 39 & 13.0 (68.8\%\up)  & 0.0041 (52.8\%\down) \\ 
\midrule
\multirow{10}{*}{ConvNet} & No-defense & 82.9 (-) & 43 & 14.0 (-) & 0.7005 (-) \\ 
 & CAFL & 76.2 (6.8\down) & 360 & 120 (757.1\%\up)  & 0.3506 (49.9\%\down) \\
 & SignSGD & 32.5 (50.4\down) & 62 & 20.2 (44.2\%\up)  & 0.0219 (96.9\%\down) \\
 & JoPEQ & 81.8 (1.1\down) & 58 & 18.7 (33.6\%\up) & 0.3503 (50.0\%\down) \\  
 & GradDrop & 82.7 (0.2\down) & 53 & 17.7 (26.4\%\up) & 0.2804 (59.9\%\down) \\
 & Soteria & 82.9 (0.0) & 92 & 29.9 (113.4\%\up) & 0.6941 ($<1.0\%$\down) \\
 & SDQ & 82.9 (0.0) & 52 & 17.3 (23.6\%\up) & 0.3503 (50.0\%\down) \\ 
 & LDP & 82.0 (0.9\down) & 52 & 17.3 (23.6\%\up) & 0.7005 (0.0\%) \\ 
 & Ours & \textbf{83.3} (\textbf{0.4}\up) & 54 & 18.0 (28.6\%\up) & 0.3503 ( 50.0\%\down) \\ 
\midrule
\multirow{10}{*}{ResNet18}
 & No-defense & 79.4 (-) & 50 & 16.3 (-)& 10.6655 (-)  \\ 
 & CAFL & 35.5 (43.9\down) &  $>2,880$ &  960 ($>5,789.6\%$\up) & 5.3328 (50.0\%\down)  \\ 
 & SignSGD & 17.1 (63.2\down) & 79 & 25.8 (58.0\%\up)& 0.3330 (96.9\%\down)  \\ 
 & JoPEQ & 78.5 (0.9\down) & 88 & 29.3 (132.5\%\up)& 5.3327 (50.1\%\down)  \\ 
 & GradDrop & 79.1 (0.3\down) & 58 & 19.3 (18.4\%\up) & 4.2262 (60.4\%\down)  \\ 
 & Soteria & 79.3 (0.1\down) & 102 & 33.3 (104.0\%\up) & 10.6626 ($<1.0\%$\down)  \\ 
 & SDQ & 78.9 (0.5\down) & 70 & 23.3 (36.8\%\up)& 5.3327 (50.1\%\down)  \\ 
 & LDP & 78.6 (0.8\down) & 63 & 22.1 (26.2\%\up) & 10.6655 (0.0\%)  \\ 
 & Ours & \textbf{79.8} (\textbf{0.4}\up) & 75 & 25.0 (53.4\%\up) & 5.3327 (50.1\%\down) \\ 
\bottomrule
\end{tabular}
\end{threeparttable}
}
\end{table}

%% file: Tables/tab_nlpbase.tex
\begin{table}[t]
\centering
\caption{Defensive schemes against DLG, TAG and LAMP attacks in NLP tasks. The dataset is CoLA and model is Base BERT.}\label{nlpbase}
\resizebox{\linewidth}{!}{
\begin{tabular}{lcccccccccccc}
\toprule
\multirow{3}{*} & 
\multicolumn{3}{c}{Non-defense} & 
\multicolumn{3}{c}{\begin{tabular}[c]{@{}c@{}}GradDrop\\($\theta_{GradDrop} = 0.2$)\end{tabular}} &
\multicolumn{3}{c}{\begin{tabular}[c]{@{}c@{}}DP\\(noise = 0.005)\end{tabular}} &
\multicolumn{3}{c}{\begin{tabular}[c]{@{}c@{}}Ours\\(p = 0.999)\end{tabular}} \\
\cmidrule(l){2-4}\cmidrule(l){5-7}\cmidrule(l){8-10}\cmidrule(l){11-13}
 & R-1 & R-2 & R-L & R-1 & R-2 & R-L & R-1 & R-2 & R-L & R-1 & R-2 & R-L \\
\midrule
DLG & 52.44 & 15.56 & 48.44 & 30.87 & 0.00 & 30.71 & 41.17 & 2.00 & 39.35 & \textbf{24.722} & \textbf{0.00} & \textbf{24.722} \\  
TAG & 70.06 & 11.22 & 47.52 & 75.111 & 30.00 & 56.89 & 41.00 & 0.00 & 37.00 & \textbf{23.528} & \textbf{0.00} & \textbf{23.278} \\ 
\textbf{LAMP} & 74.51 & 28.53 & 59.33 & 82.22 & 27.85 & 66.10 & 32.67 & 4.00 & 32.67 & \textbf{19.78} & \textbf{0.00} & \textbf{19.78} \\  
\bottomrule
\end{tabular}
}
\end{table}

%% file: Tables/tab_defense.tex
\begin{table*}[t]
\sethlcolor{green} 
\centering
\caption{Examples reconstructed by DLG, TAG, and LAMP attacks. The reconstructed terms that align with ground truth are highlighted.}
\resizebox{0.9\linewidth}{!}{
\begin{tabular}{p{2cm}p{6cm}p{6cm}p{6cm}}
\toprule
 & \textbf{DLG} & \textbf{TAG} & \textbf{LAMP}  \\
\midrule
Reference 
 & [CLS] who do you think that will question seamus first? [SEP] 
 & [CLS] who do you think that will question seamus first? [SEP]
 & [CLS] who do you think that will question seamus first? [SEP] \\
\midrule
Non-defense 
 & [CLS] out \hl{who will question seamus}ythe center? \hl{you} rather [SEP]
 & [CLS] hoping \hl{first} might \hl{will} \hl{seamus question} \hl{think} \hl{who} \hl{?} \hl{you} [SEP]
 & [CLS] \hl{seamus} would \hl{who} \hl{seamus} which \hl{will} what \hl{you} \hl{question?} [SEP] \\
\midrule
GradDrop
 & [CLS] robthing certainly air linux pga factor specifically \hl{seamus} selected [SEP]
 & [CLS] \hl{seamus questions} \hl{who} \hl{think} / registration mind \hl{you} \hl{?} \hl{who} [SEP]
 & [CLS] \hl{do you think} \hl{who} \hl{will} \hl{?} \hl{seamus} \hl{first} \hl{question} \hl{seamus} [SEP] \\
\midrule
DP
 & [CLS] robthing certainly air linux pga factor specifically \hl{seamus} selected [SEP]
 & [CLS] just\hl{?} someone \hl{seamus}tor trim five romeo puzzle also [SEP]
 & [CLS] perhapspartisan donateants \hl{seamus} carbon bounty jennie tidebaum [SEP] \\
\midrule
Ours
 & [CLS] if hit vision tank [MASK] ky titular speed like car [SEP]
 & [CLS] tony dna just flames spiritsgirl avery season - famed [SEP]
 & [CLS] [PAD] [PAD] [PAD] [PAD] [PAD] [PAD] [PAD] [PAD] [PAD] [PAD] [SEP] \\
\bottomrule
\end{tabular}}
\label{defense2}
\end{table*}

%% file: Tables/tab_reconstruction_from_flipping.tex
\begin{table}[t]
\centering
\caption{Defence performance against reconstruction attacks that imputes flipped bits (outliers) with gradient mean or 0.}
\label{tab_reconstruction}
\resizebox{0.4\linewidth}{!}{
\begin{tabular}{cccc}
\toprule
\multirow{2}{*}{$p$} & \multirow{2}{*}{\begin{tabular}[c]{@{}c@{}}Defence\\Performance\end{tabular}} & \multicolumn{2}{c}{Impute outliers to} \\ \cmidrule(l){3-4} 
 &  & Mean & 0 \\ \midrule
\multirow{3}{*}{0.98} & PSNR & 9.428 & 9.357 \\
 & SSIM & -0.082 & -0.076 \\
 & LPIPS & 0.723 & 0.730 \\
\midrule
\multirow{3}{*}{0.95} & PSNR & 7.636 & 7.673 \\
 & SSIM & -0.092 & -0.113 \\
 & LPIPS & 0.759 & 0.751 \\
\midrule
\multirow{3}{*}{0.90} & PSNR & 6.637 & 6.664 \\
 & SSIM & -0.127 & -0.126 \\
 & LPIPS & 0.751 & 0.757 \\ 
\bottomrule 
\end{tabular}
}
\end{table}

%% file: Tables/tab_various_epsilons.tex

\begin{table}[t]
\centering
\caption{Model accuracy and defense performance with various $\epsilon$ values. Number in \zehang{orange} denotes LDP in same setting cannot converge. }
\label{tab_various_epsilons}
\resizebox{0.8\linewidth}{!}{
\begin{tabular}{@{}cccccc@{}}
\toprule
\multirow{2}{*}{$\epsilon$} & \multirow{2}{*}{$p$} & \multirow{2}{*}{Accuracy (\%)} & \multicolumn{3}{l}{Defense Performance} \\ \cmidrule(l){4-6} 
 &  &  & PSNR & SSIM & LPIPS \\ \midrule
\zehang{1.38} & \zehang{0.80} & \zehang{77.7} & \zehang{4.555} & \zehang{0.070} & \zehang{0.734}\\
2.20 & 0.90 & 79.2 & 4.600 & 0.079 & 0.736 \\
2.94 & 0.95 & 79.4 & 4.606 & 0.094 & 0.733 \\
3.89 & 0.98 & 79.8 & 5.114 & 0.190 & 0.732 \\ \bottomrule
\end{tabular}
}
\end{table}